\documentclass[12pt]{iopart}
\usepackage{graphicx}
\usepackage{subfigure}
\usepackage{dcolumn}
\usepackage{bm}
\newcommand{\IGNORE}[1]{}
\newcommand{\R}     {\mathcal{R}}
\newcommand{\A}     {\bf A}
\newcommand{\B}     {\bf B}
\newcommand{\C}     {\bf C}
\newcommand{\D}     {\bf D}

\newcommand{\Lor}  {\mathcal{L}}

\newcommand{\invunits}  {\textrm{nm} / \textrm{meV}}
\newcommand{\minvunits} {\textrm{nm} / \mu\textrm{eV}}

\begin{document}

\title[Resonant carrier dynamics]{Resonant carrier dynamics in strongly biassed superlattices}
\author{Pavel Abumov$^{1,2}$ and D. W. L. Sprung$^1$}
\address{$^1$ Department of Physics and Astronomy, McMaster University, Hamilton, Ontario L8S 4M1 Canada}
\address{$^2$ Department of Electrical Engineering, University of California, Santa Cruz 98800 CA}
\ead{pabumov@ucsc.edu}

% \date{\today}

\begin{abstract}
We study coherent electron dynamics in a biassed undriven ideal
semiconductor superlattice coupled to the continuum, near energy
level anticrossings. In particular, we examine the dependence of
wavepacket dynamical characteristics on electric field detuning, and
investigate mixed regimes involving a superposition of energy level
anticrossings showing both Rabi oscillations and resonant
tunnelling. In earlier work~[Phys. Rev. B {\bf 75} 165421 (2007)],
Rabi and Zener resonances were shown to have a common origin, and a
criteria for the occurrence of either was proposed. The present
results allow a better understanding of the nature of an
interminiband resonance, which can be useful in the areas of
microwave radiation generation and matter manipulation on the
particle level, as well as demonstrate an alternative approach to
examining electron level structure of a finite superlattice.
\end{abstract}

\pacs{73.23.-b, 73.21.Cd, 78.20.Bh, 78.30.Fs}

%\keywords{quantum transport, Rabi oscillations, resonant Zener tunnelling, semiconductor superlattice, interminiband dynamics}
\submitto{\JPCM}

\maketitle

\section{\label{sec:intro}Introduction}

Carrier dynamics in a biassed superlattice (SL) has remained an
active topic during the last two decades. Knowledge of the
underlying physical processes is necessary for a better
understanding of perpendicular transport of carriers in
multi-quantum well systems, as well as for successful development of
applications, such as microwave radiation
generation~\cite{Bloch_Oscillator, Bloch_oscillator3,
Bloch_Oscillator2, Shimada04}, quantum computing~\cite{Zrenner02,
Wang05} and matter manipulation on the particle
level~\cite{Beam_Splitter}.

Studies focussed on coherent carrier dynamics have revealed some
dynamical features of great interest (see, for
example,~\cite{Latge05} and references therein) arising from the
interplay between interminiband oscillations and tunnelling to the
continuum. These studies contribute to understanding of the famous
quantum-mechanical problem of tunnelling in presence of
dissipation~\cite{Jagdeep99, Beham03, Vasanelli02} that has been
considered for superlattices in \cite{Europhysics05}.

Whereas typically such studies involve a number of approximations,
we adopted a computational approach with few limitations  beyond those
implicit in the model of a single electron in a biassed periodic potential.
This enabled us to consider resonant dynamics at high and moderate bias
in a finite superlattice, specifically Rabi oscillations (RO) and
resonant tunnelling (RT), and to establish a relation between these
two fundamental types of interminiband transport by simple means. A
carrier dynamics solution relying on wavepacket time evolution also
brings out certain features that are difficult to calculate
otherwise, such as carrier behaviour in the vicinity of a resonance,
and dependence of the period of Rabi oscillations on resonance
index.

The main purpose of this paper is to further analyze and explain the
results obtained using the methods of our earlier
work~\cite{Abumov07}. In particular, our approach helps to study
near-resonance behaviour in more detail and understand the damping
mechanism of RO and ways to reduce it. It also contributes to a link
between RO and RT, an instance where quantum transport theory has
been lacking to date. These results can be applied to any system
possessing a Wannier-Stark ladder structure including photonic
crystals~\cite{Photonics4} and optical lattices~\cite{Arimondo07}.

\section{\label{sec:res}Resonant interminiband dynamics}

This work focusses on longitudinal motion of a single electron in a
zero-temperature biassed superlattice. We solve
the time-dependent Schr\"odinger equation (TDSE) for $\Psi(x,t)$  in
a biassed periodic potential
    \begin{eqnarray}       %eq01
         V(x) &=& \sum_{n=-\infty}^{\infty} V_{SL}(x-n d) + xF~,    \qquad {\rm where}   \nonumber \\
    V_{SL}(x) &=& \frac{V_0}{2} \, \Big[ \tanh \frac{x+a/2}{\sigma}
                                    - \tanh \frac{x-a/2}{\sigma} \, \Big] \Theta(x) \, \Theta(d-x)~.
%                        && \qquad \times \quad \Theta(x) \, \Theta(d-x)~.
    \label{eq:01}
    \end{eqnarray}
where $\Theta(x)$ is the Heaviside function vanishing for $x < 0$,
$F=-eE$ is a uniform bias, $E$ is the electric field, and
$V_{SL}(x)$ is the model superlattice potential in a unit cell of
the periodic system, with width $d$ and barrier thickness $a$.

We consider coherent electron transport and thus omit electron
scattering and relaxation processes; numerous studies of carrier
coherency limits have been done in the past~\cite{Coherency1,
Coherency2}. These destructive processes are weak enough at low
temperatures, to allow up to 14 Rabi oscillations to
occur~\cite{Roskos92}.

\begin{figure}                      %fig00
      \leavevmode
      \begin{center}
        \includegraphics [height=3.5cm,angle=270,keepaspectratio=true]{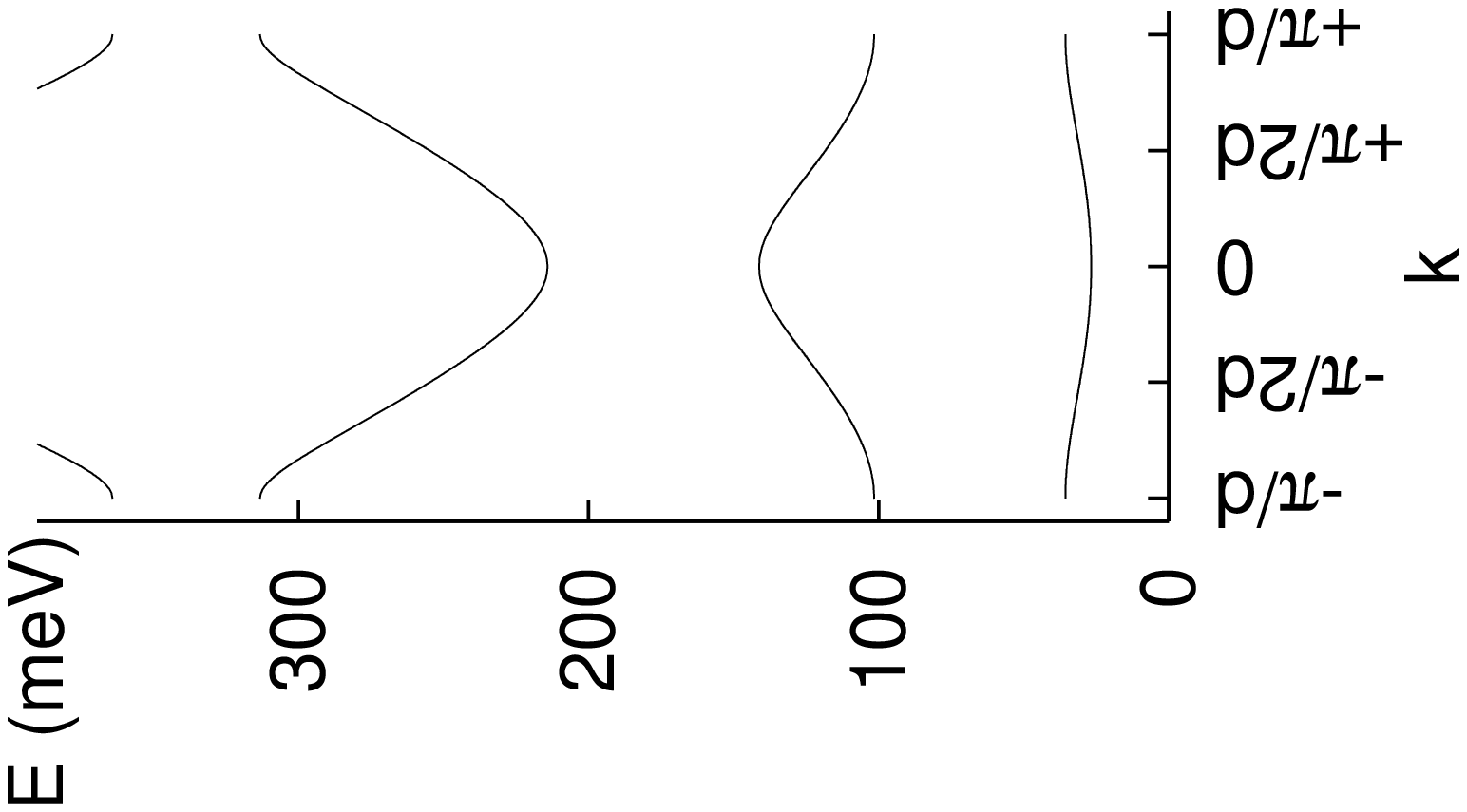}
        \includegraphics [height=3.5cm,angle=270,keepaspectratio=true]{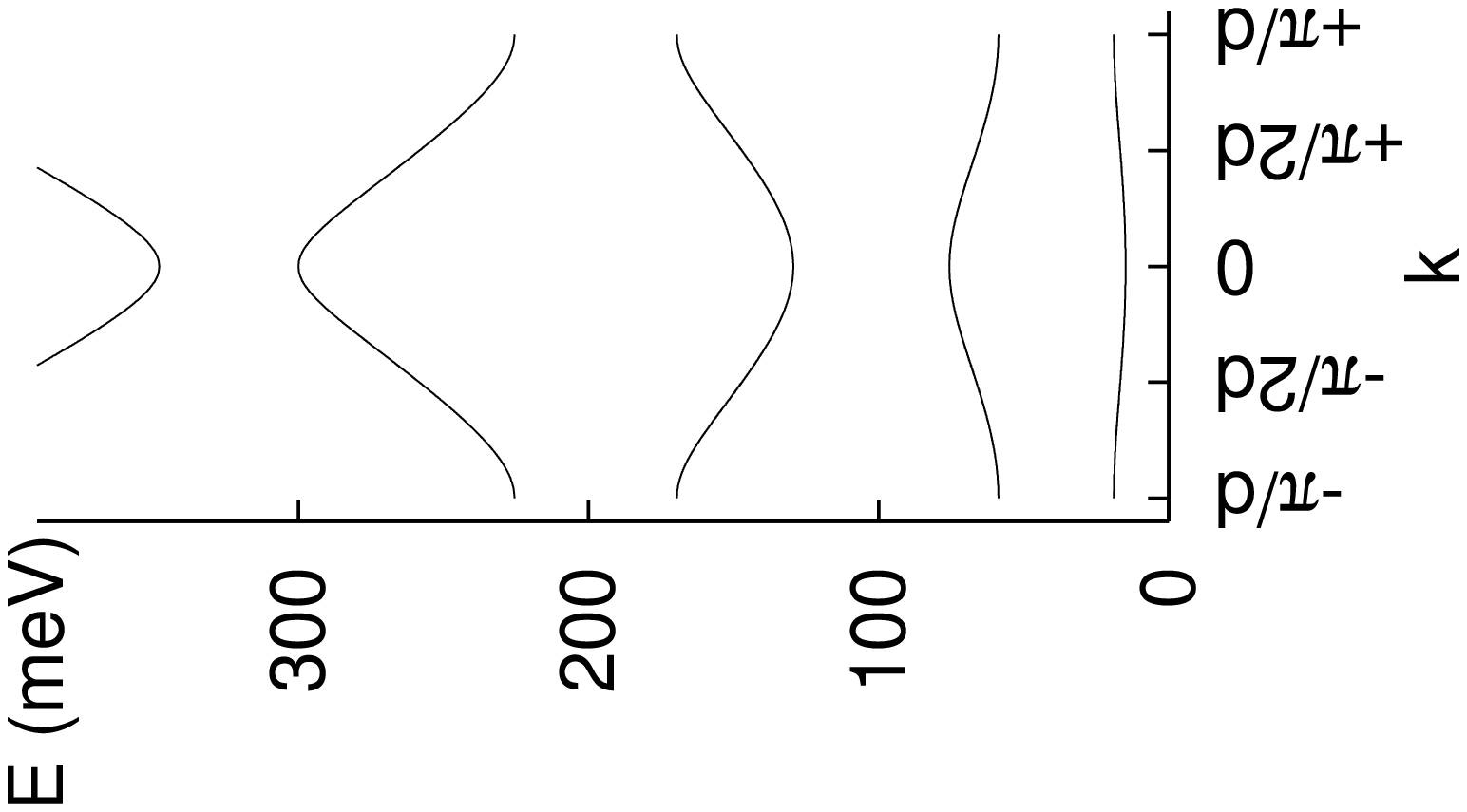}
        \includegraphics [height=3.5cm,angle=270,keepaspectratio=true]{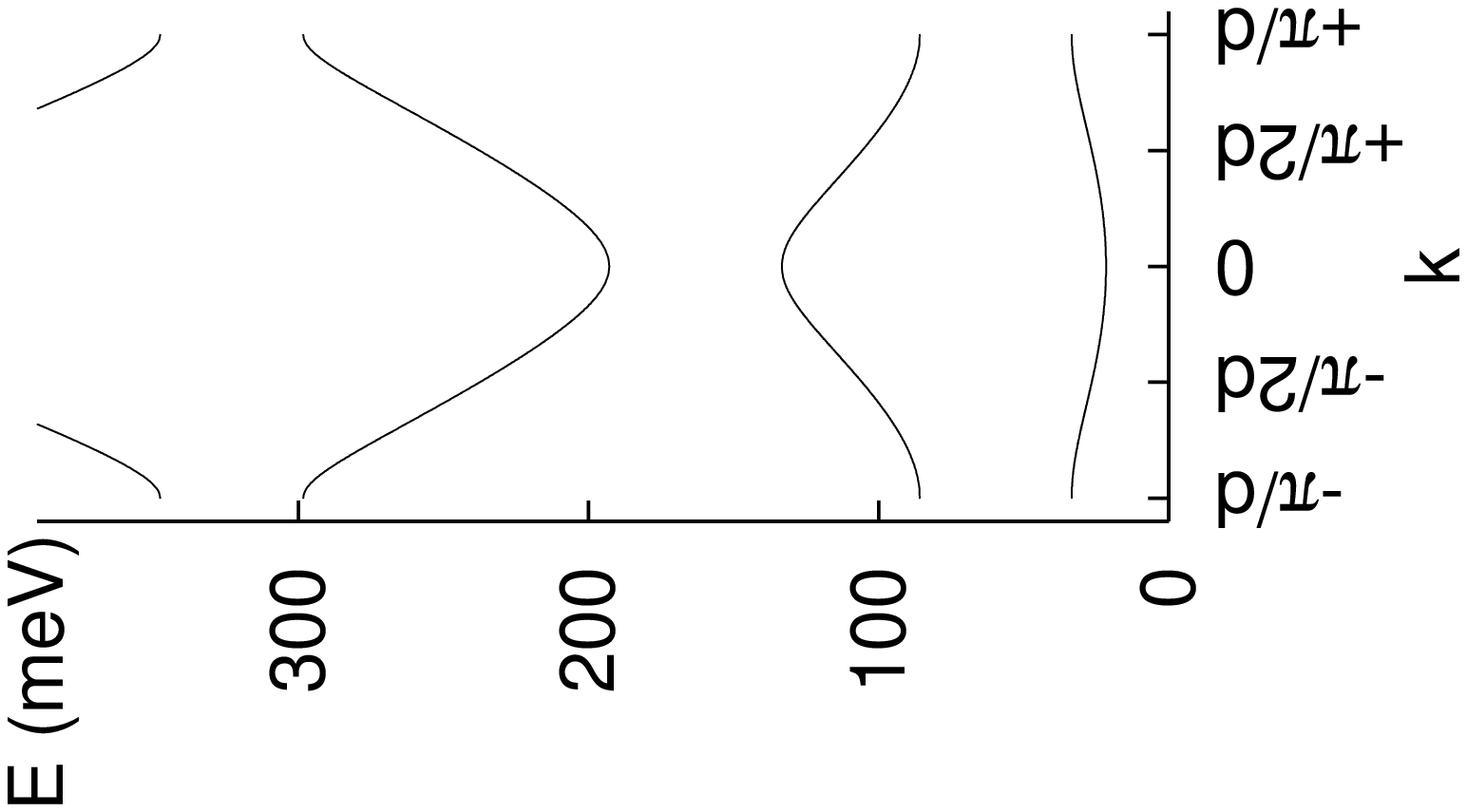}
        \includegraphics [height=3.5cm,angle=270,keepaspectratio=true]{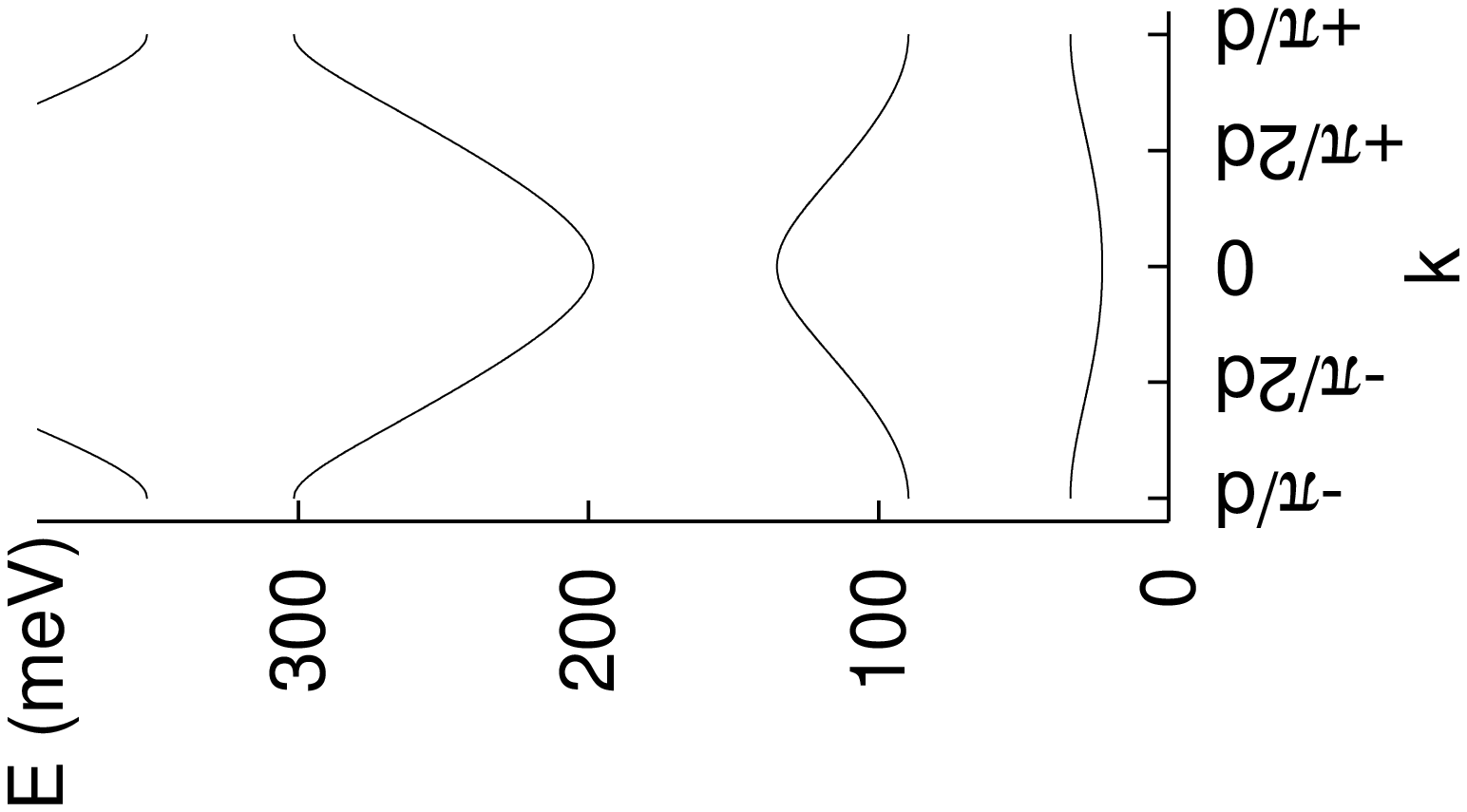}
      \end{center}
      \caption[]{Band structure of the superlattice potentials in absence of bias
                (from left to right: samples A, B, C and D).}
      \label{fig:00}
    \end{figure}

 Several GaAs/Ga$_{1-x}$Al$_{x}$As layered heterostructures
labelled $X = A,\, B,\, C,\, D$ are considered, whose parameters are
shown in Table~\ref{tbl:01} and their band structure in
Figure~\ref{fig:00}. The layer profile function $V_{SL}$ avoids
discontinuities in the potential, which allows easier programming of
the solver, and is more realistic than the often used square barrier
approximation. Our numerical solutions of Equation~\ref{eq:01} use
discrete transparent boundary conditions~\cite{Moyer}, which reduces
the size of the space domain in which we must operate. For details,
see reference~\cite{Abumov07}.

The norm of the quasibound part of the wavepacket $\rho$ and miniband
occupancy $\rho_{\nu}$ are convenient properties by which to monitor
interminiband dynamics; miniband occupancy is the wavepacket
projection onto a tight-binding miniband~$\nu$:
$$\rho_{\nu}=\sum_k|\langle\,\Psi(x,t)\,|\,W_{\nu}^k(x)\,\rangle|^2~.$$
Here $W_{\nu}^k(x)$ stands for a Wannier-Stark (WS) quasibound state
corresponding to energy level $E_{\nu}^k$ centered on the well with
index~$k$, and belonging to miniband $\nu$ ($\nu$=1,2,~\ldots); our
initial wavepacket is centered on the well with index~0. The
tight-binding Wannier functions from miniband $\nu$ are denoted as
$w_{\nu}(x)$.

To refer to an interminiband resonance originating from an
anticrossing of energy levels $E_{\nu}^k$ and $E_{\mu}^{k+n}$ in
biassed sample~X, we will use the symbol $\R_{\nu\mu}^n(X)$ and we
will denote the resonant bias (the value of bias at which the peak
of a resonance is observed) as $F_n$, $n$ being the resonance index
($n$ = 1,2,~\ldots). When of little importance, some indices may be
omitted for brevity. For convenience the symbol $G$ will stand for
inverse bias $1/F$ ($G_n = 1/F_n$). Unless specified otherwise, time
is measured in units of the Bloch period $T_B={2\pi\hbar}/{Fd}$.

It should be noted that computing a complete complex-energy
spectrum for a biassed multilevel system is a significant
task beyond the aims of the paper, as can be seen from
previous works devoted to this problem (for example, the formalism
in~\cite{Gluck_review, Gluck1} utilizes existence of poles of the
system scattering matrix). We instead consider a simplified
framework proven suitable for the purpose of studying transient
transport processes.

    \subsection{\label{sec:res:model}Rabi oscillations model}

Typically, Rabi Oscillations (RO) are a result of interminiband
transitions under external radiation when its frequency approaches
the system's intrinsic value $\omega_{12}
=({E_2-E_1})/{\hbar}$~\cite{Akulin}. This pumping provides
excitation of carriers and their subsequent spontaneous emission of
photons, which is widely applied in quantum cascade
lasers~\cite{first_QC_laser}. A comprehensive overview of RO in the
two-miniband approximation has been given by Ferreira and
Bastard~\cite{Ferreira97} and further mathematical details can be
found in references~\cite{Brandi05,Rotvig94,Gluck5}. In a biassed
SL, interminiband transitions may also occur in the absence of
external radiation~\cite{Ferreira94, Bastard94, Voisin97}: when
energy levels from coupled Wannier-Stark ladders (WSL) align in
neighbouring wells, a carrier can easily tunnel between them.

    \begin{table}[tl]       %tbl01
      \begin{center}
            \begin{tabular}{ccccc}                                              \hline \hline
              Name   & $V_0$,\ meV& $d$, nm(ML) & $a$, nm(ML) & $\sigma$, nm \\ \hline
            Sample A &  212       &  13.0 (46)  &  3.1 (11)   &   0.4        \\
            Sample B &  250       &  17.3 (61)  &  2.5 (9)    &   0.4        \\
            Sample C &  212       &  13.0 (46)  &  2.3 (8)    &   0.4        \\
            Sample D &  212       &  13.0 (46)  &  2.5 (9)    &   0.4        \\ \hline \hline
            \end{tabular}
      \end{center}
      \caption[]{Geometric parameters of the model potentials used in simulations.
                Barrier height of 212~meV corresponds to $x$=0.18 and of 250~meV to
                $x$=0.3 in the GaAs/Ga$_{1-x}$Al$_{x}$As structure; ML
                stands for monolayer thickness, 1 ML = 0.283 nm.}
      \label{tbl:01}
    \end{table}

For an undriven SL ({\it i.e.} not exposed to an external
oscillating electric field) a simplified yet fruitful analogy to an
unbiassed system under external radiation can be made. For
illustration, consider Bloch oscillations of an electron in a
biassed SL, where the electron oscillates in a Bloch domain, with
turning points of the motion defined by the edges of a miniband. One
can think of this process not as a motion in space, but as
oscillation in the local kinetic energy between the upper and lower
band edges. For example, let an electron at $t=0$ have the mean
position $\langle x_0 \rangle$ and the fixed initial energy
$E_0=(E_U+E_L)/2$, half way between upper and lower edges, and let a
positive bias $F$ be applied in the positive $x$-direction. An
electron moving towards larger values of $x$ will eventually 
collide with the lower band edge $E_L$, as
the latter rises with slope $F$. The amplitude of motion in space is
$L/2 = (E_0-E_L)/F = (E_U - E_L)/(2F) \equiv (\delta E/2)/F$, where
$\delta E$ is the miniband width and $L$ is the Bloch Oscillation
(BO) domain width. The two domains, space and energy, are directly
related through the bias value $F$.

Assuming very large carrier lifetime $1/\Gamma$ compared to a
typical oscillation period for moderate bias, and taking the WS
states to be stationary and orthogonal, we can apply the
acceleration theorem $\hbar\, ({d\vec{k}}/{dt})=\vec{F}$, combined
with the dispersion relation, to describe wavepacket evolution over
a Bloch oscillation cycle. Making use of the decomposition
$V(x)=V_{SL}+Fx$ we can write
    \begin{eqnarray}       %eq02
      \langle\, x(t)\, \rangle &=& \frac{1}{\hbar} \int_{0}^{t} \,
      \frac{dE(k_0+{Ft}/{\hbar})}{dk} \: dt
    \label{eq:02}
    \end{eqnarray}
When the lowest minibands are tightly bound (which is common for a
semiconductor SL), carrier dispersion  in MB$n$ can be approximated
as $E(k)=E_n+ ({\delta E}/2) \cos(kd+n\pi)$ and the mean position
evolves as  $\langle x(t)\rangle-\langle x(t_0) \rangle = (L/2)
\sin\Big( \omega_B (t-t_0)\Big)$, where $\omega_B$ is the Bloch
angular frequency.

Thus, the local kinetic energy is now time-dependent:
    \begin{eqnarray}
    E\,(t - t_0) &=& (E_U + E_L)/2 - F \Big(\langle x(t)\rangle - \langle x(t_0)\rangle\Big)     \nonumber \\
             &=& E_0 - ( F L/2)\, \sin\Big( \omega_B (t-t_0)\Big)                        \nonumber \\
             &=& E_0 - (\delta E/2)\, \sin\Big( \omega_B (t-t_0)\Big)~.                    %\nonumber
    \end{eqnarray}
The wavepacket's evolution in coordinate space at fixed energy can 
effectively be replaced by its evolution in kinetic energy space.
With minor variations, the same viewpoint can be adopted for 
intrawell oscillations in a similar fashion. Note that the 
intrawell oscillation frequency is a natural frequency of the
system: $\omega_{nm}=({E_m-E_n})/{\hbar}$.

Given the fact that interminiband transitions of an electron are a
result of quantum interference of Bloch and intrawell
oscillations~\cite{Abumov07}, one can extend this analogy to Rabi
oscillations. Then the SL would be subject to a field of natural
frequency $\omega_{nm}$, modulated by the Bloch frequency $\omega_B$.
In presence of the $\omega_{nm}$ harmonic, this system is
poised to undergo an interminiband transition, with a certain Rabi
frequency $\omega_R$. This is similar to the well-understood system of
a two-level atom under an external monochromatic radiation resonance
field $f_n$. The Schr\"odinger equation for such a system transforms
into
    \begin{eqnarray}       %eq03
     - i\hbar\,\frac{\partial\Psi(x,t)}{\partial t} \,=\, (\hat{H_0}+\hat{V}) \, \Psi(x,t) \nonumber\\
       {\rm with}\ \ \  \hat{V} \,=\, \hat{x}\, f_n \,\cos(\omega_R t),
    \label{eq:03}
    \end{eqnarray}
where $\hat{H_0}$ refers to an undriven biassed SL in the
tight-binding approximation, $\hat{x}$ is the dipole transition
operator between two energy levels (minibands in our case), and
$f_n$ is the amplitude of the external electric field.

The factor $\hat{x}f_n$, interpreted as a transition operator
in our effective model, represents the strength of
interminiband coupling and depends on many factors, including the
applied bias and the potential shape. For a small $\hat{x}f_n$,
perturbation theory can be applied and an expression for the
population of the second level evolving from an initial wavepacket
$\Psi(x,0)=W_1(x)$ can then be obtained in the same fashion as for
an irradiated two-level atom. In our notation, for a two-level
system driven off its ground state~\cite{Akulin} one obtains:
    \begin{eqnarray}       %eq04
         \frac{\rho_2}{\rho}(t) &=& \Big(\frac{\rho_2}{\rho}\Big)^{max}\Lor(G)
            \,\sin^2 \frac{\pi t}{T_R^{max}  \sqrt{\Lor(G)}}   \nonumber\\
         \frac{\rho_1}{\rho}(t) &=& 1 - \frac{\rho_2}{\rho}(t), \qquad {\rm with} \nonumber\\
          \Lor(G) &\equiv& \bigg[1 + \Big(\big(G-G_n\big) / \Gamma\Big)^2
                     \bigg]^{-1}   \quad  {\rm where} \nonumber\\
      \Gamma &=& {x_{0n}}/{(E_2-E_1)}~, \nonumber\\
      T_R^{max} &=& {d}/{x_{0n}}~, \nonumber\\
      x_{nm} &=& \langle\, W_1^n(x)\,|\,x\,|\,W_2^m(x)\, \rangle~,
    \label{eq:04}
    \end{eqnarray}
with $n$ being the index of the resonance considered. These
equations are straightforwardly extended to an arbitrary pair of
interacting minibands, subject to validity of the perturbation
theory used. It is interesting that in order to analytically predict
the $\rho_2/\rho\,(t)$ curve in the entire near-resonance region we
require only the values of the dipole matrix element $x_{nm}$ and of
$\big(\rho_2/\rho\big)_{max}$, computed at the resonant bias field.

This simplistic derivation for BO in a two-level tightly
bound system applies to the wavepacket's `centre of mass'. We
neglected any change in dispersion of the wavepacket in the process of
BO and thus it is only an approximation. Since we consider moderate to
high fields in this paper, carrier decay is rapid enough to reduce
the influence of this factor. As will be seen in subsequent
sections, this simple model provides a surprisingly good fit to the
simulation data.

In the case of a strong field, the domain width in space of the
wavepacket's `center-of-mass' BO is typically smaller than a
potential cell width. This implies that: (i) energy levels are
sparse due to large splitting of WSL; hence near a resonance there
is one preferred tunnelling path between MB1 and MB2; (ii) at a
resonance, WS states are reasonably localized and have an
exponentially vanishing tail, so we expect that $x_{0n} \propto
e^{-n}$ (which has been predicted by a two-level atom
model~\cite{Bastard94} and also has been explicitly calculated for a
driven multiband SL~\cite{Gluck_review}).

The above argument assumed the system to be in a steady state.
However, in practice an initial non-equilibrium configuration
undergoes a relaxation process to a state with lower
potential energy.
We studied the evolution of an initial wavefunction $\Psi(x,t=0)=c_1 w_1(x)+c_2
w_2(x)$, a linear combination  of tight-binding Wannier
states. (Obtaining correct quasibound WS functions is not important
for our primary goal to study steady-state RO dynamics.) The
steady-state tunnelling process showed little dependence on
the linear combination: components of $\Psi(x)$ which
belong to higher minibands tunnel out rapidly during the initial
relaxation period, which only scales down the norm of the quasibound
wavepacket being observed. The coefficients $c_1$
and $c_2$ should be set so that after an initial period of
relaxation of a non-equilibrium state, the resulting RO are clear
and of sufficient magnitude. The maximum magnitude of RO between two
lowest minibands reaches nearly unity for the two extreme cases
$|c_1|=1$ and $|c_2|=1$, with less carrier amplitude decay for the
former choice. Hence, we adopted the initial wavefunction
$\Psi(x, t=0)=w_1(x)$ for most of our results.

\begin{figure}[b]         %fig01
    \leavevmode
    \begin{center}
        \includegraphics [height=3.7cm,angle=0]{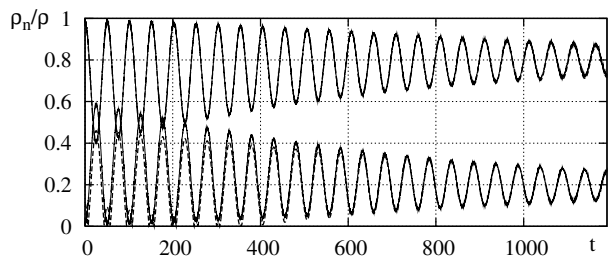}
    \end{center}
    \caption[]{Occupancy functions (solid lines) and their fit using
                Equation~\protect \ref{eq:05}
                (broken lines almost completely hidden behind them) 
               for $\Psi(x,0)=w_1(x)$ near a resonance $\R_{12}^3(\A)$
                at $G=0.44058\ \invunits$ (0.8 HWHM away from the resonant bias);
                the upper curve corresponds to $n=1$ and the lower to $n=2$.}
    \label{fig:01}
\end{figure}

\subsection{\label{sec:res:curves}Near-resonance behaviour}

\begin{figure}         %fig02
    \leavevmode
    \begin{center}
    \begin{tabular}{cc}
        {\includegraphics[height=4cm,angle=270,keepaspectratio=true]{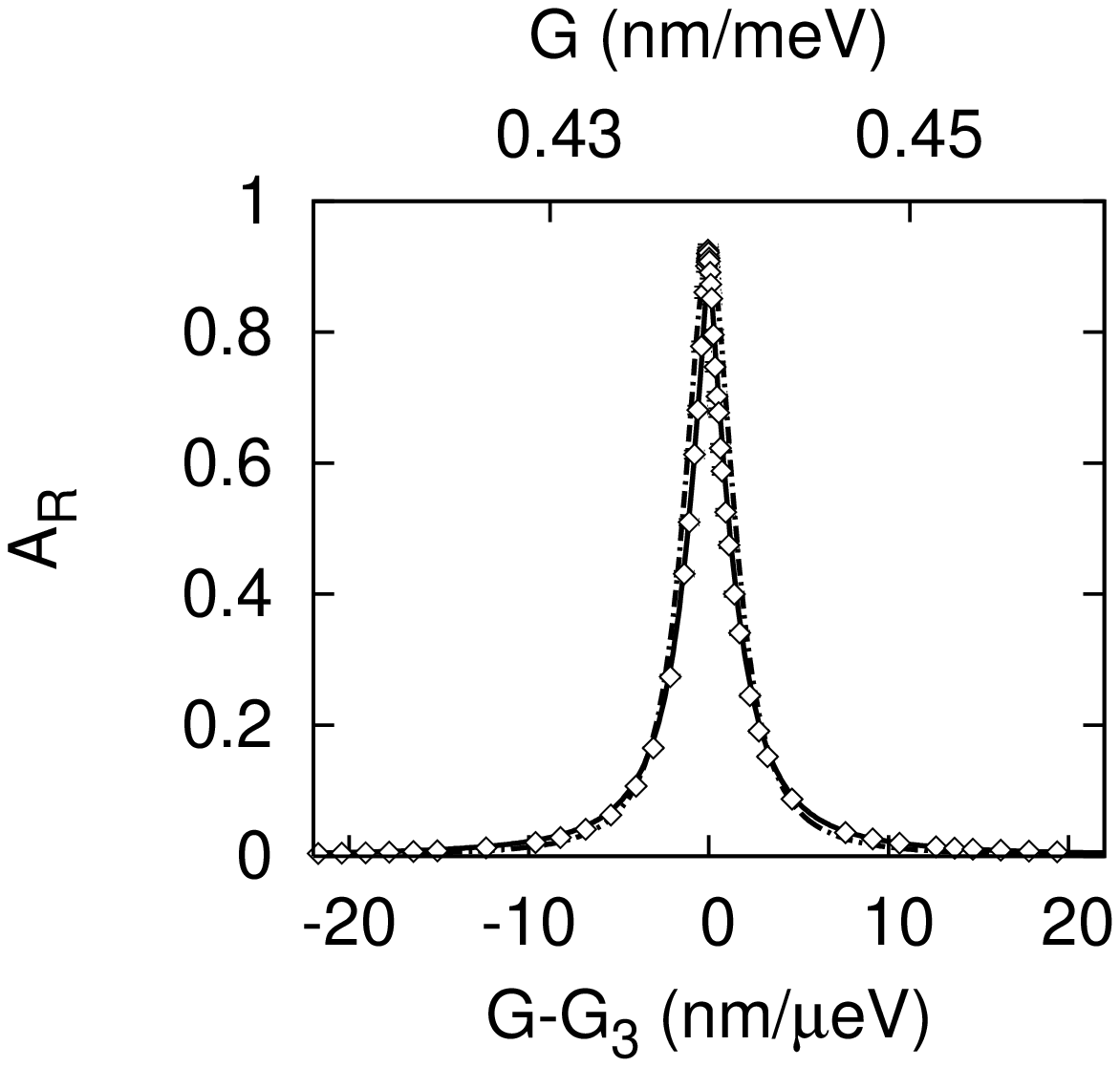}} &
        {\includegraphics[height=4cm,angle=270,keepaspectratio=true]{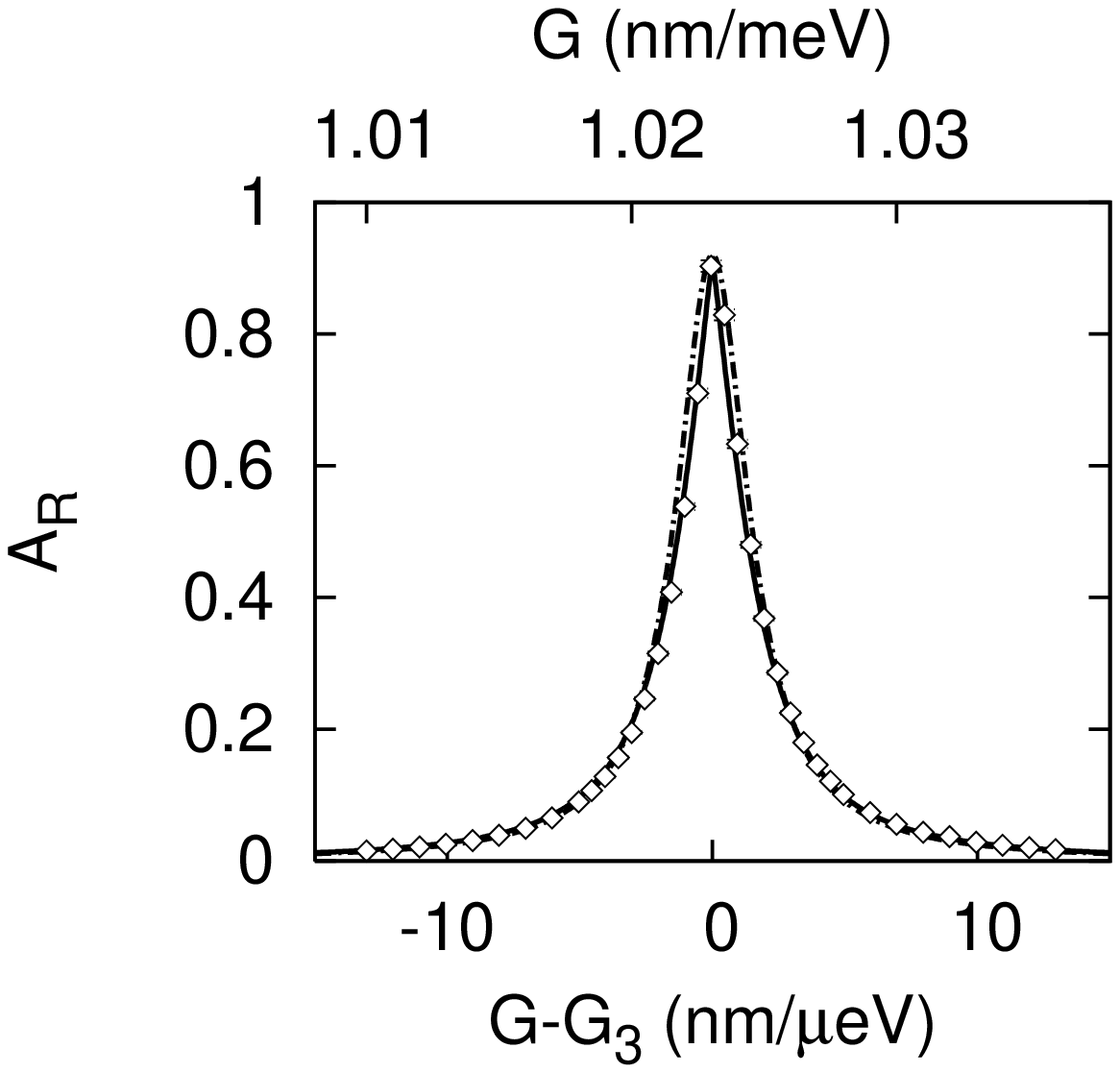}} \\

        {\includegraphics[height=4cm,angle=270,keepaspectratio=true]{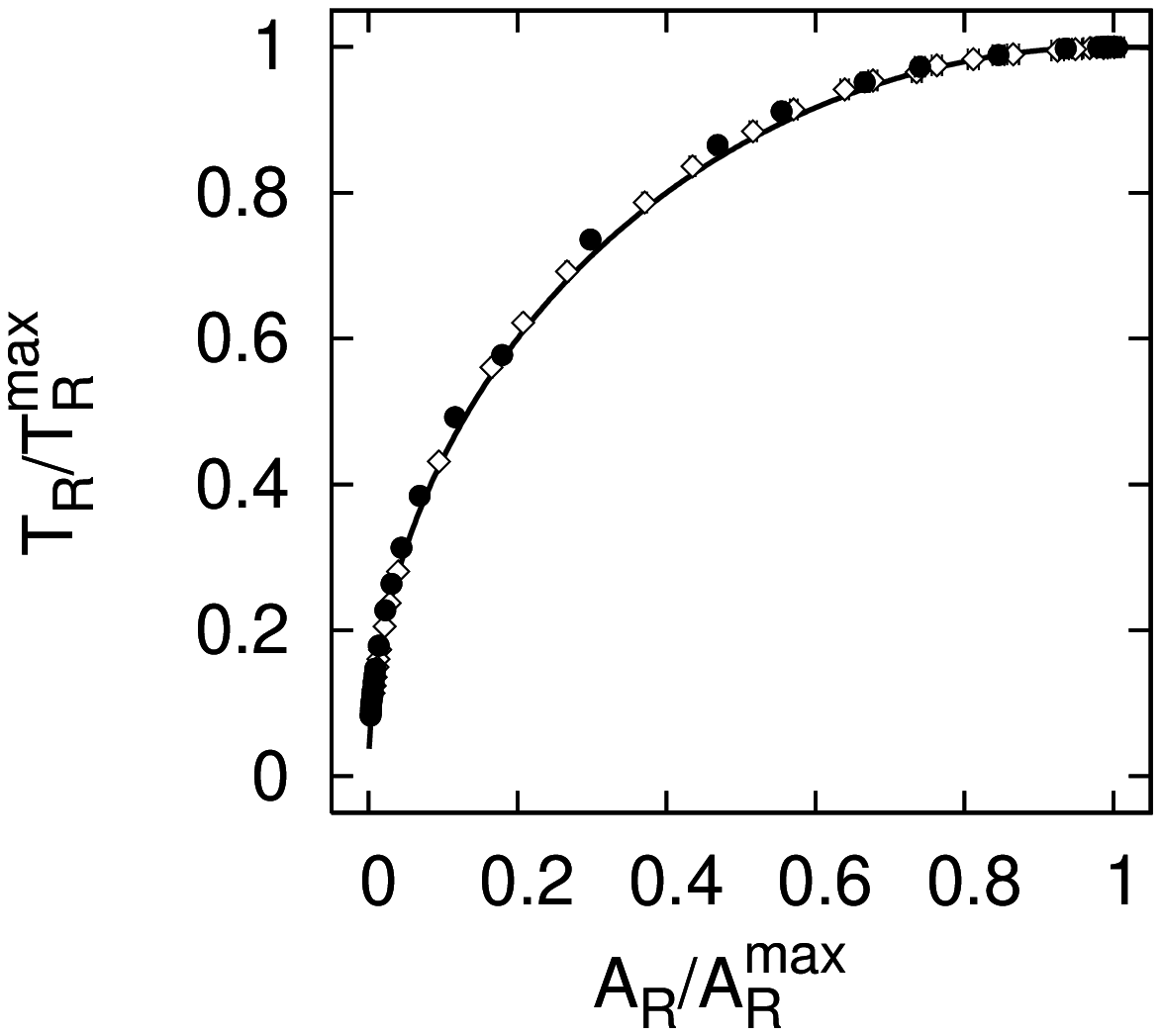}} &
        {\includegraphics[height=4cm,angle=270,keepaspectratio=true]{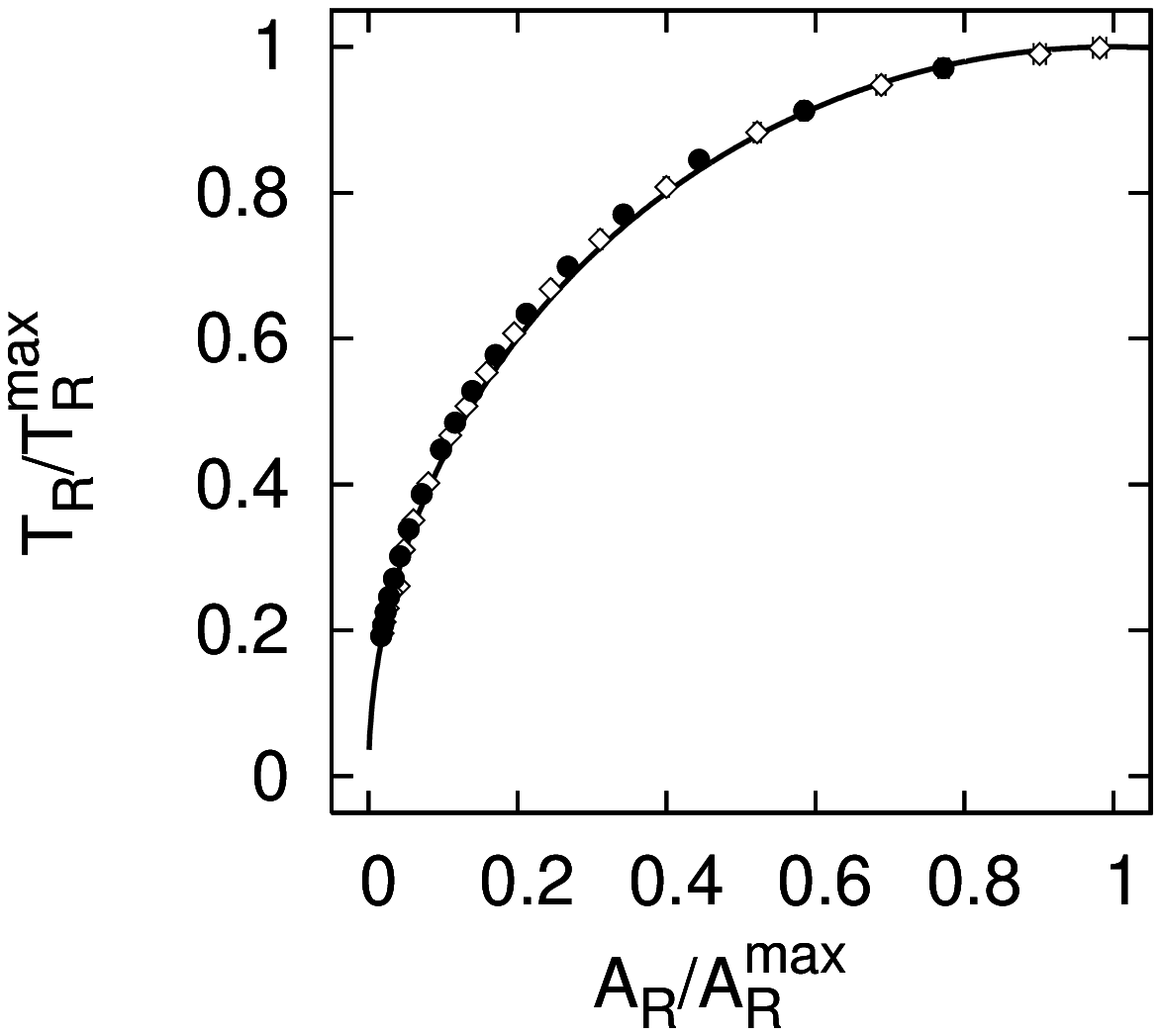}} \\

        {\includegraphics[height=4cm,angle=270,keepaspectratio=true]{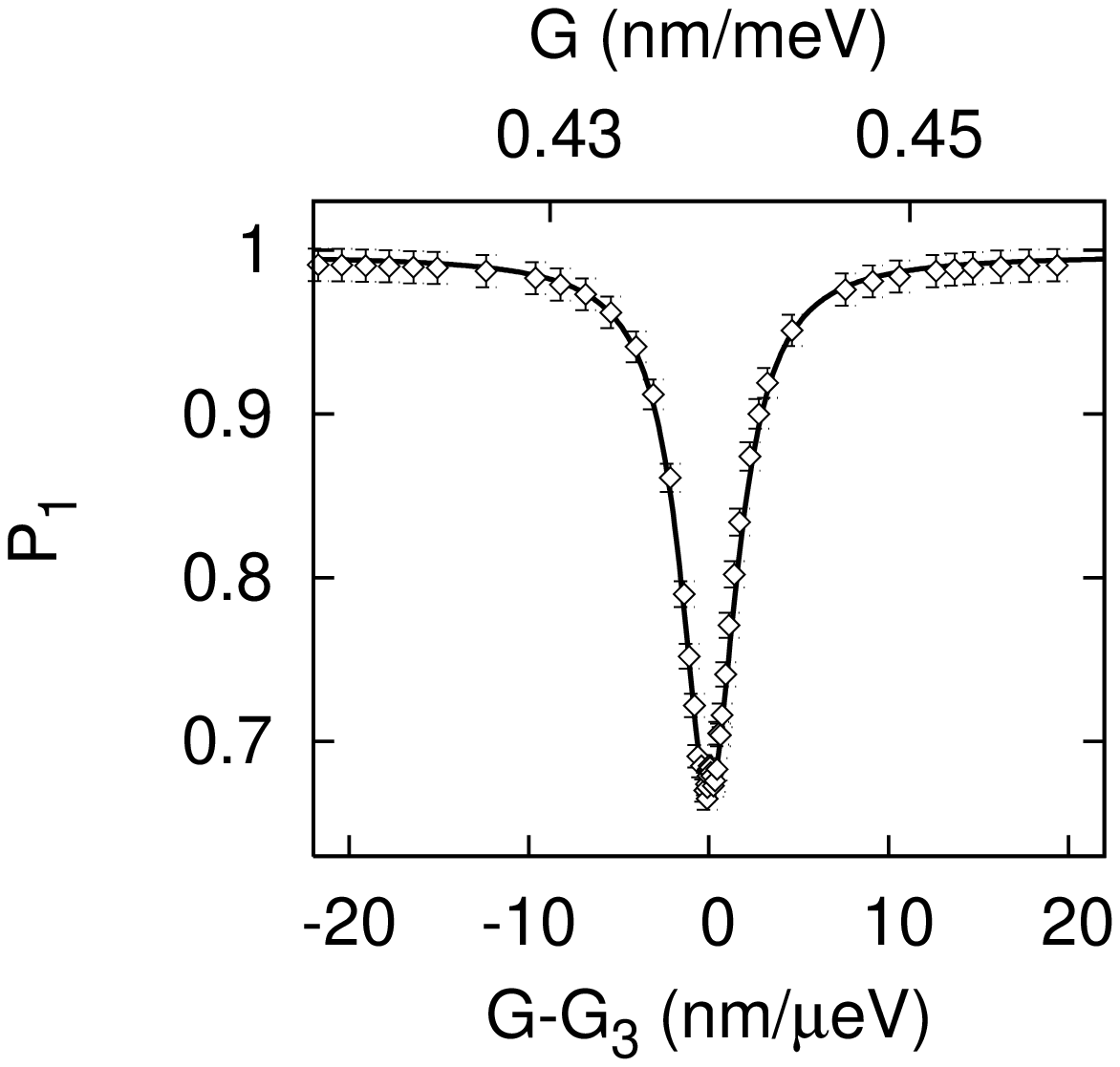}} &
        {\includegraphics[height=4cm,angle=270,keepaspectratio=true]{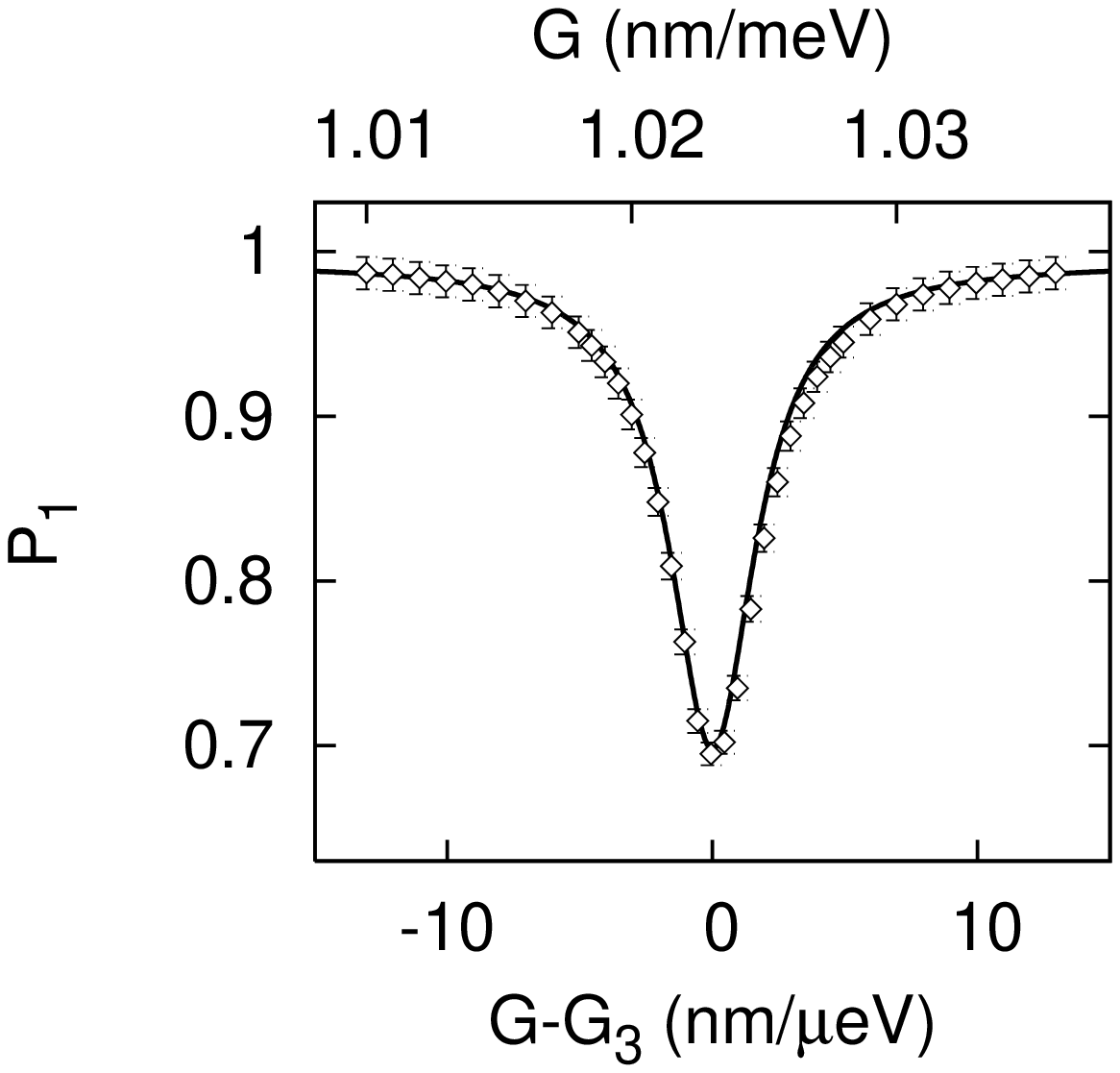}} \\

        {\includegraphics[height=4cm,angle=270,keepaspectratio=true]{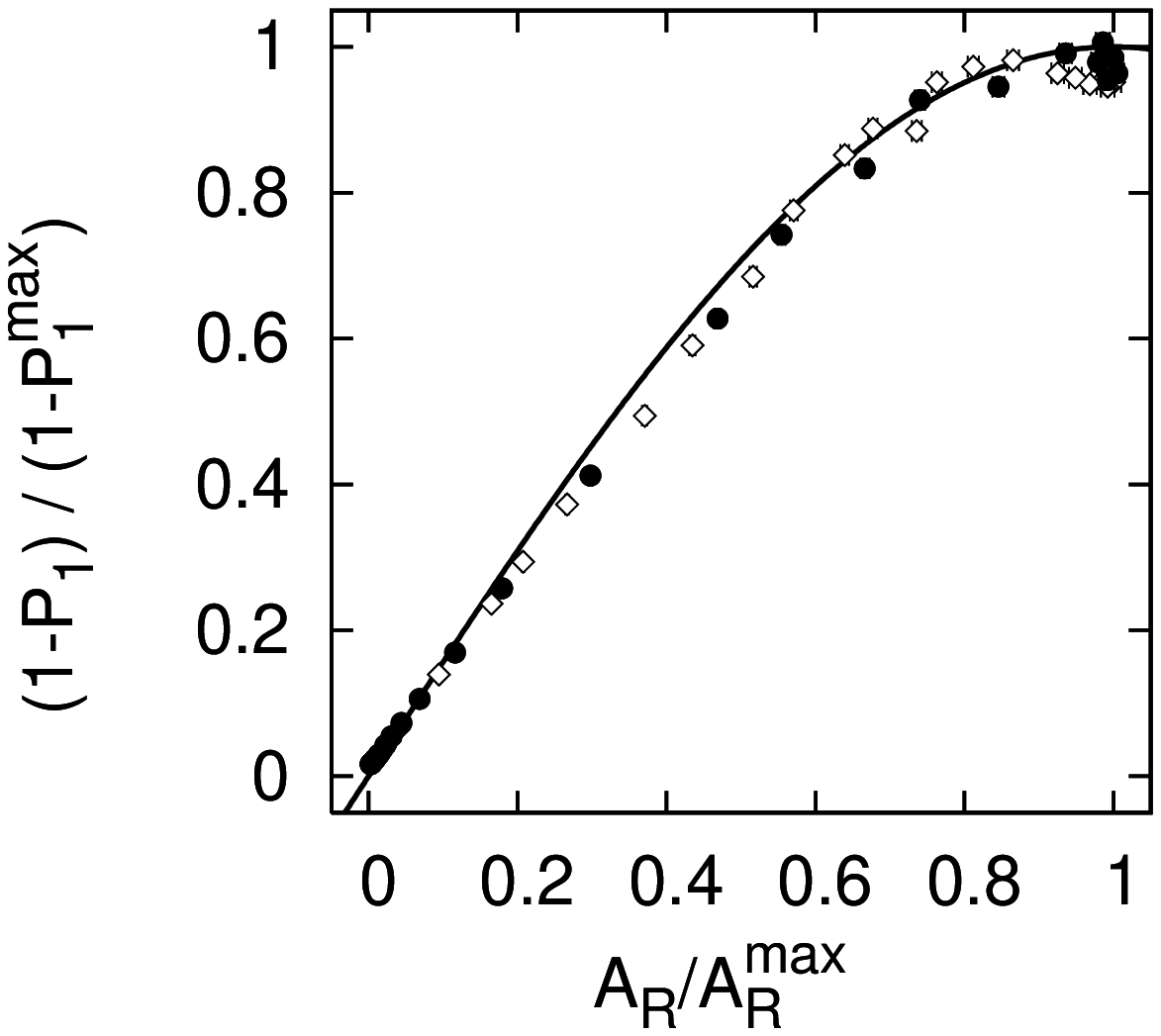}} &
        {\includegraphics[height=4cm,angle=270,keepaspectratio=true]{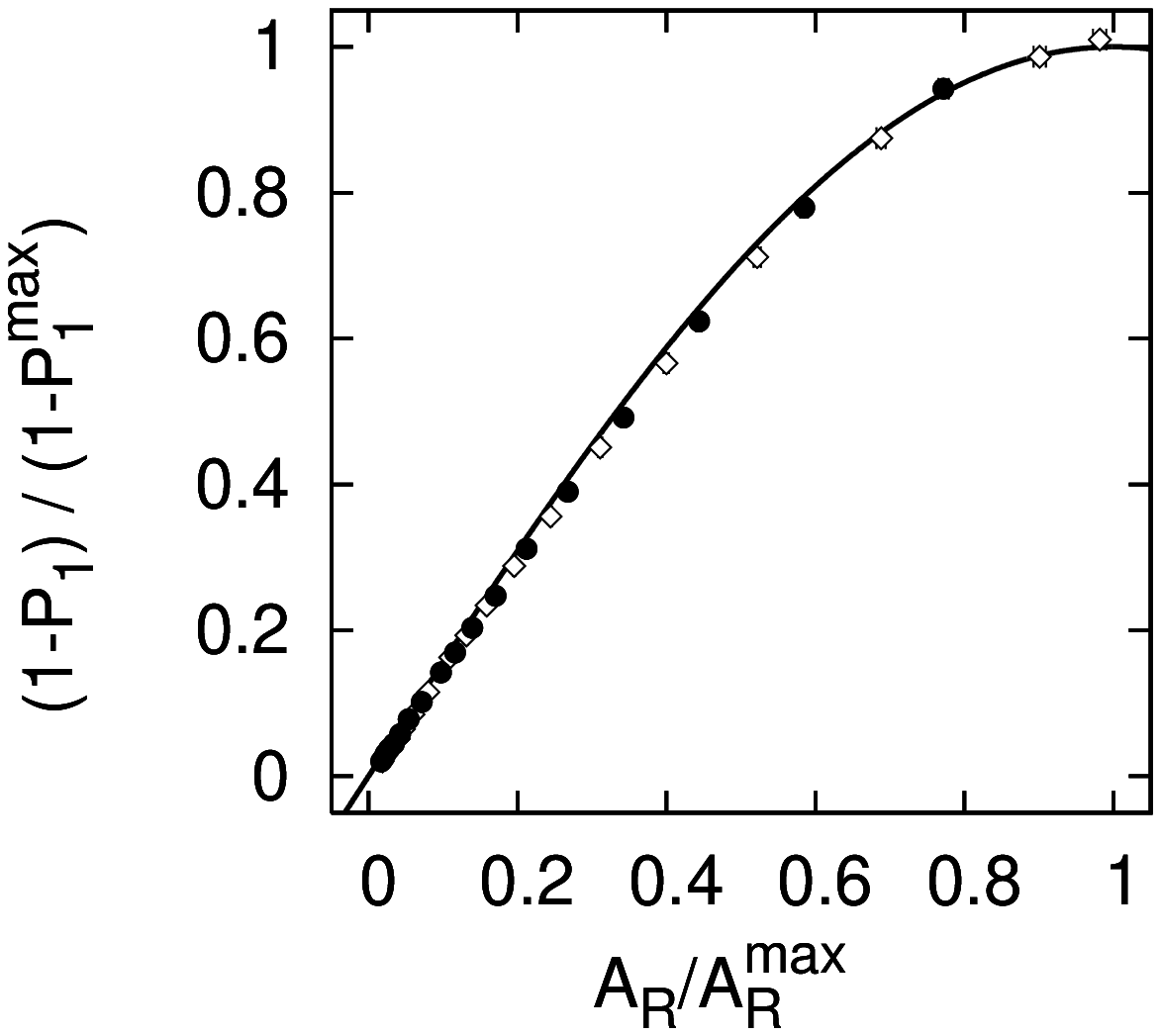}} \\

        {\includegraphics[height=4cm,angle=270,keepaspectratio=true]{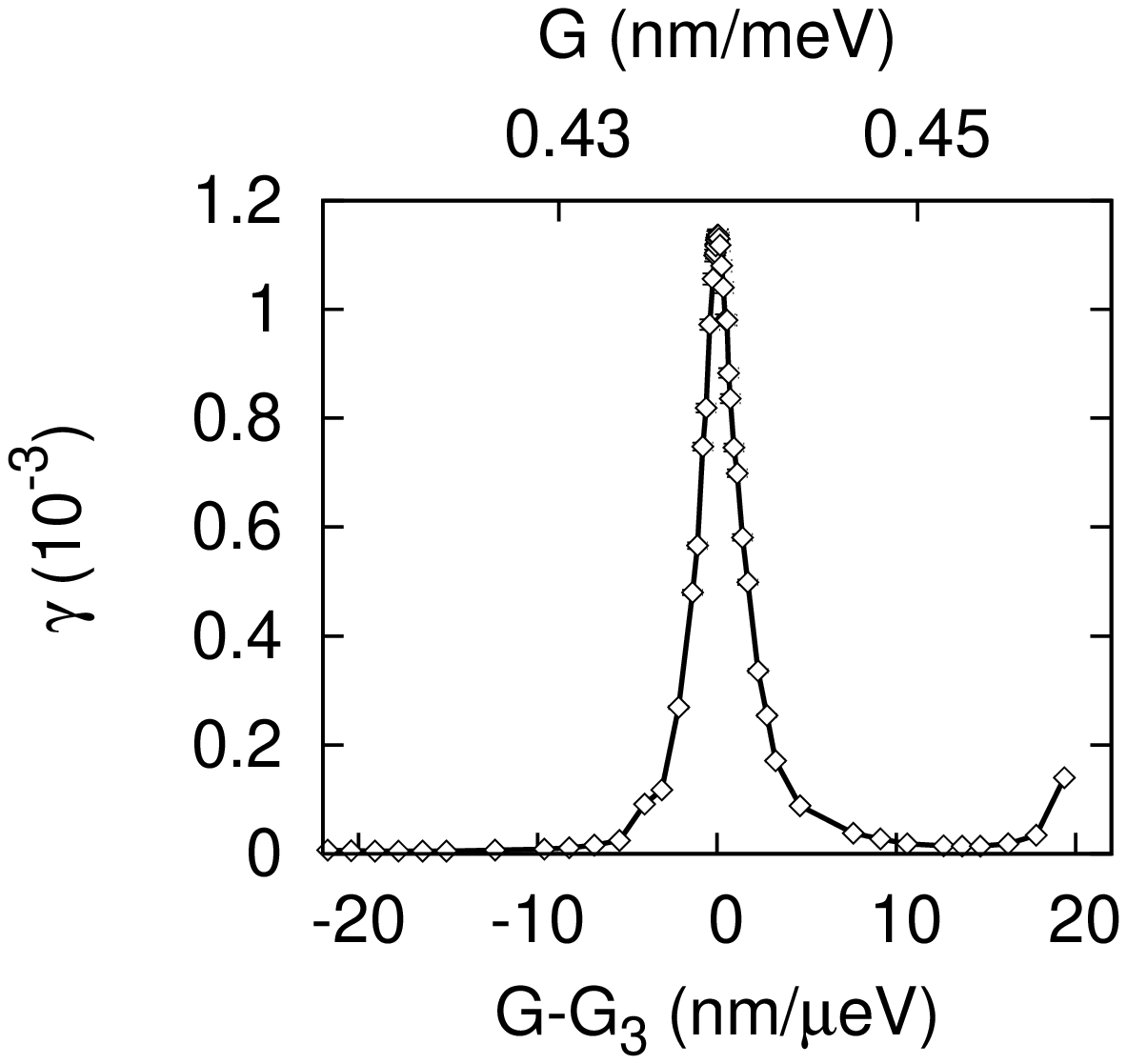}} &
        {\includegraphics[height=4cm,angle=270,keepaspectratio=true]{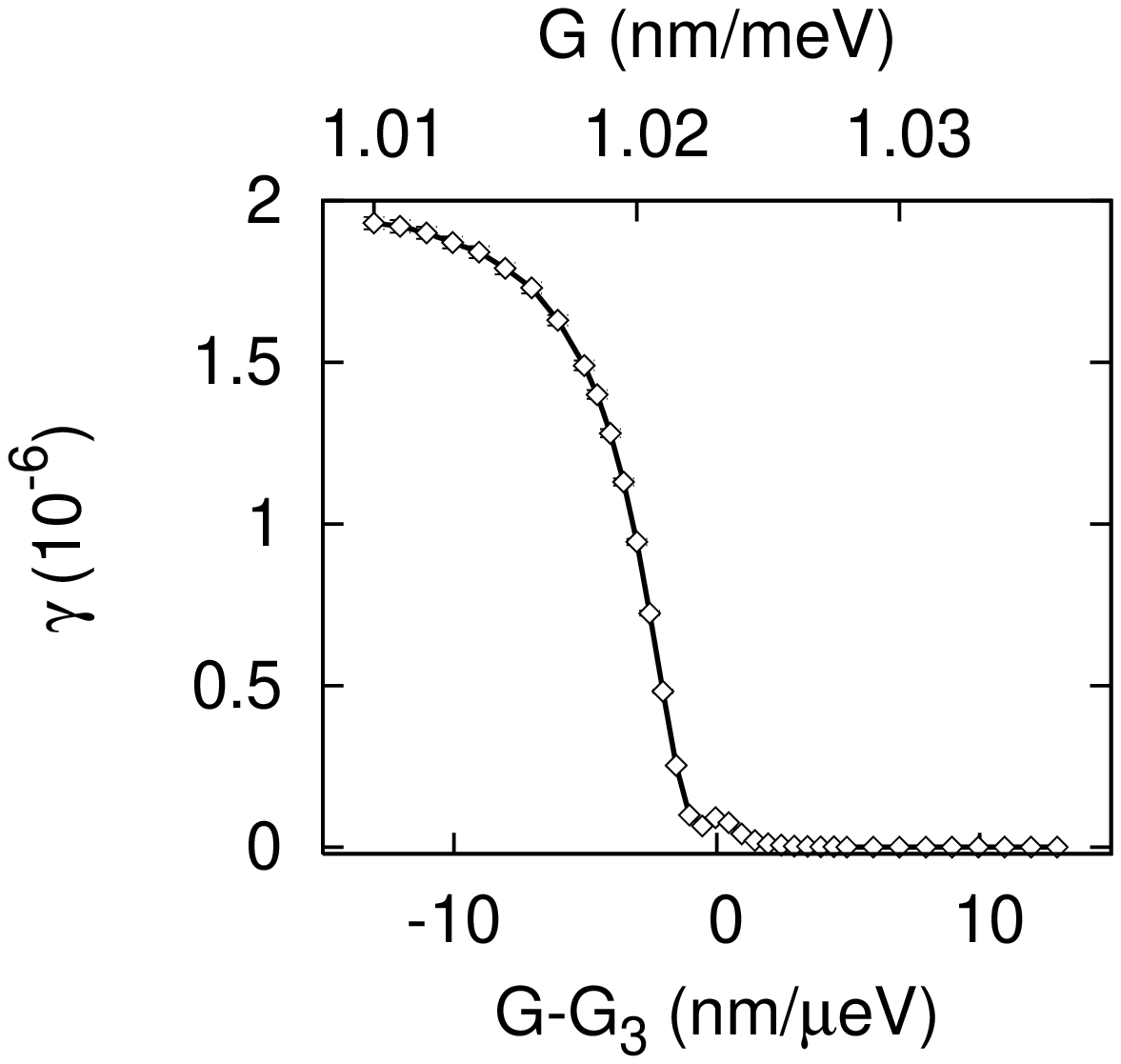}} \\

    \end{tabular}
    \caption[]{Behaviour of key dynamical parameters for $\R_{12}^3(\A)$ (left column)
                and $\R_{12}^3(\B)$ (right column).
                Top panel shows RO amplitude vs. inverse bias (chain-dotted line shows a
                Lorentzian fit $\Lor(G)$, solid line shows fit using Equation~\protect\ref{eq:07a}),
                second from top panel: relative RO amplitude vs. relative RO period
                (filled circles correspond to $G<G_3$, empty circles to $G>G_3$; solid line shows
                fit using Equation~\protect\ref{eq:07a} given Equation~\protect\ref{eq:06}),
                third from top panel: asymptotic occupancy of the first miniband vs.
                inverse bias (solid line shows fit using Equation~\protect\ref{eq:07}),
                fourth from top panel: relative RO period vs. asymptotic occupancy of
                the first miniband (filled circles correspond to $G<G_3$, empty circles
                to $G>G_3$; solid line shows fit using Equation~\protect\ref{eq:07});
                bottom panel: carrier decay rate vs. inverse bias.}
    \label{fig:02}
    \end{center}
\end{figure}

In a Wannier-Stark ladder, energy level anticrossings show a
dramatic increase in carrier tunnelling rate, which is ascribed to
resonant Zener tunnelling. When studied in detail, carrier dynamics
at some of these anticrossings exhibits a prominent
oscillatory pattern (Rabi oscillations) which decays rapidly.
For clarity we will call the latter Rabi resonances and the
others tunnelling resonances. While both are anticrossings of
complex energy levels (complex due to their finite width), it will
allow us to differentiate the two types. At strong fields,
above-barrier tunnelling resonances start to occur; we will include
them as tunnelling resonances also.

A study of isolated Rabi resonances in~\cite{Abumov07} revealed that
a carrier undergoing RO produces damped oscillations in miniband
occupancy, of the form:
    \begin{eqnarray}       %eq05
\rho_n \big{/} \rho\,(t) &=& P_n + A_R\,e^{-
\gamma_At}\cos\,\Big(\,\frac{2\pi t}{T_R}\,+\phi_n \Big)~, \nonumber \\
    {\rm with} \ \ \rho\,(t) &=& \exp \Big( -\gamma t - \sigma\,
            (1-e^{-\gamma_{ne} t})\;\Big)
    \label{eq:05}
    \end{eqnarray}
in agreement with the general theory~\cite{Vasko99}. Here $T_R$,
$A_R$ and $\gamma_A$ are the period, amplitude and decay rate of
Rabi oscillations respectively; $\phi_n$ is an  initial phase
determined by a particular form of $\Psi(x,0)$ ({\it e.g.} $\phi_1
\approx 0$ for $\Psi(x,0)=w_1(x)$); $\gamma$ is the decay rate of
the entire wavepacket and $P_n$ is the asymptotic value of
$\rho_n/\rho\,(t)$ in the limit $t\rightarrow\infty$. The term
$-\sigma\,(1-e^{-\gamma_{ne}t}) $ represents a dip in the decay rate
due to two-exponential decay~\cite{twoExp89} and was found to be
vanishingly small close to the resonant bias. Its presence is due to
the initial relaxation of wavepacket components orthogonal to
eigenfunctions of the current miniband, with $\sigma$ being
proportional to the norm of the orthogonal part and $\gamma_{ne}$
being the relaxation rate, typically larger by two orders of
magnitude  than $\gamma$. This term compensates for the change in
states' orthogonality at higher fields. A typical example of time
evolution of near-resonant interminiband occupancy dynamics, and its
fit using Equation~\ref{eq:05} is shown in Figure~\ref{fig:01}.

    \begin{figure}         %fig03
      \leavevmode
      \begin{center}
        \includegraphics[height=4cm,angle=270,keepaspectratio=true]{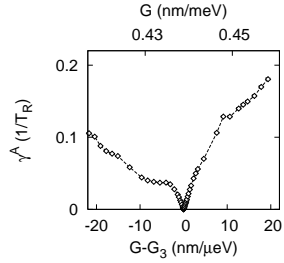}
        \caption[]{Damping rate of RO vs. inverse bias for $\R_{12}^3(\A)$, in
         units of $1/T_R$.}
      \label{fig:03}
      \end{center}
    \end{figure}

Since the parameters in Equation~\ref{eq:05} provide a good
description of carrier dynamics for moderate bias, and we will
investigate their behaviour near a resonance. It has been found
previously~\cite{Abumov07} that the period $T_R$ of Rabi
oscillations shows a root-Lorentzian peak around the resonant bias:
    \begin{eqnarray}       %eq06
      T_R(G) &=& T_R^{max} \, \sqrt{\Lor(G)}~,
      \label{eq:06}
    \end{eqnarray}
with the Lorentzian $\Lor$ defined in Equation~\ref{eq:04}. We will
refer to the parameter $\Gamma$ of this Lorentzian as the HWHM (half-width
at half-maximum) of the resonance under consideration. It has also
been shown that the peak period of Rabi oscillations changes
exponentially with resonance index $n$:
$$T_n^{max} = T_1^{max} \Big( T_2^{max} / T_1^{max} \Big)^{n-1}~.$$

For a double quantum well system, perturbation theory predicts that
the frequency of Rabi oscillations strictly at resonant bias is
given by $\hbar \omega_R\,/2 =\, \langle\,\Psi_L\,|\,V(x)\,|\,\Psi_R
\rangle$, the tunnelling matrix element through the barrier
separating the two wells~\cite{Vasko99} and is essentially the
splitting between the energy levels. Thus the inverse of $T_R^{max}$
of Equation~\ref{eq:06} is the minimum energy level splitting at
their anticrossing.

According to these earlier findings, the occurrence of resonant bias
values also correlates well with complex energy spectrum
anticrossings. As an example, the bias values for anticrossings
between two lowest minibands in sample A, obtained from our
simulation ($G_1$=(6.9$\pm$0.2)~$\invunits$,
$G_2$=(3.4$\pm$0.2)~$\invunits$ and
$G_3$=(2.33$\pm$0.05)~$\invunits$), are reasonably close to those
calculated by K. Hino {\it et~al}.~\cite{Toshima05} (7.2~$\invunits$,
3.6~$\invunits$ and 2.4~$\invunits$, respectively). This agreement
demonstrates that interminiband motion prevails at complex energy
anticrossings.

The parameters $A_R$ and $P_n$ derived from simulation results
also have clear extrema. The following equations fit the
data quite well (see Figure~\ref{fig:02}):
    \begin{eqnarray}       %eq07
    A_R &=& A_R^{max} \bigg[1-\sqrt{1-\Lor_{\nu\mu}^n(G)}\bigg],  \\
      \label{eq:07a}
   \tilde{P}_n &=& \tilde{P}_n^{max} \sin\frac{\pi A_R}{2 A_R^{max}}, \nonumber
      \label{eq:07}
    \end{eqnarray}
with $A_R^{max}$ and $\tilde{P}_n^{max}$ being the peak values;
$\tilde{P}_n=P_n$ for the lower resonantly coupled MB and $1-P_n$
for the upper one. The other two parameters, $\gamma$ and
$\gamma_A$, are extremely sensitive to coupling to higher minibands
and their bias detuning dependence varies from one resonance to
another. However, at a resonant bias, $\gamma_A$ always reached its
virtually zero minimum, and $\gamma$ its maximum
(see Figures~\ref{fig:02},~\ref{fig:03}).

    \begin{figure}[t]         %fig04
      \leavevmode
      \begin{center}
        \includegraphics[height=4cm,angle=270,keepaspectratio=true]{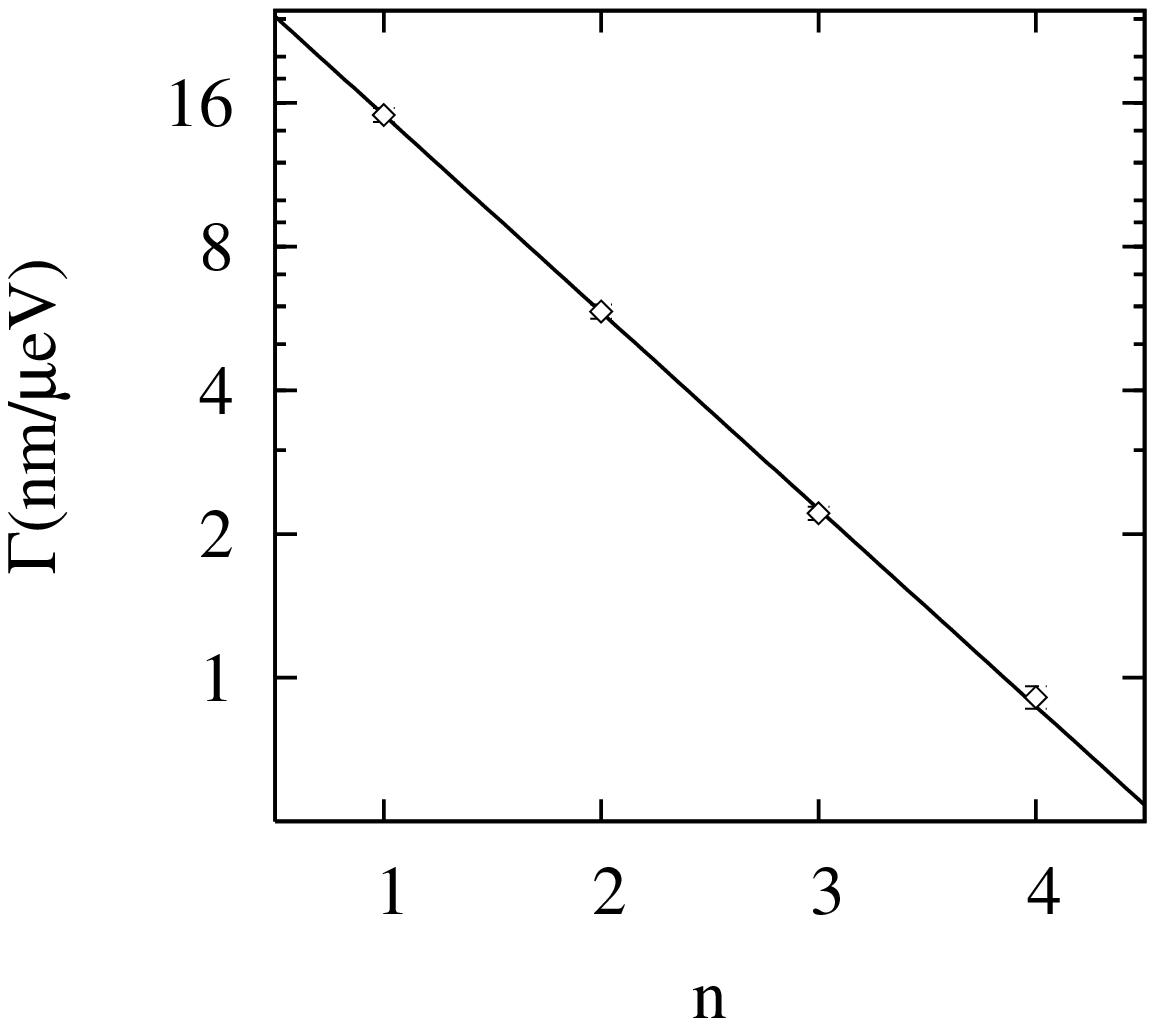}
        \includegraphics[height=4cm,angle=270,keepaspectratio=true]{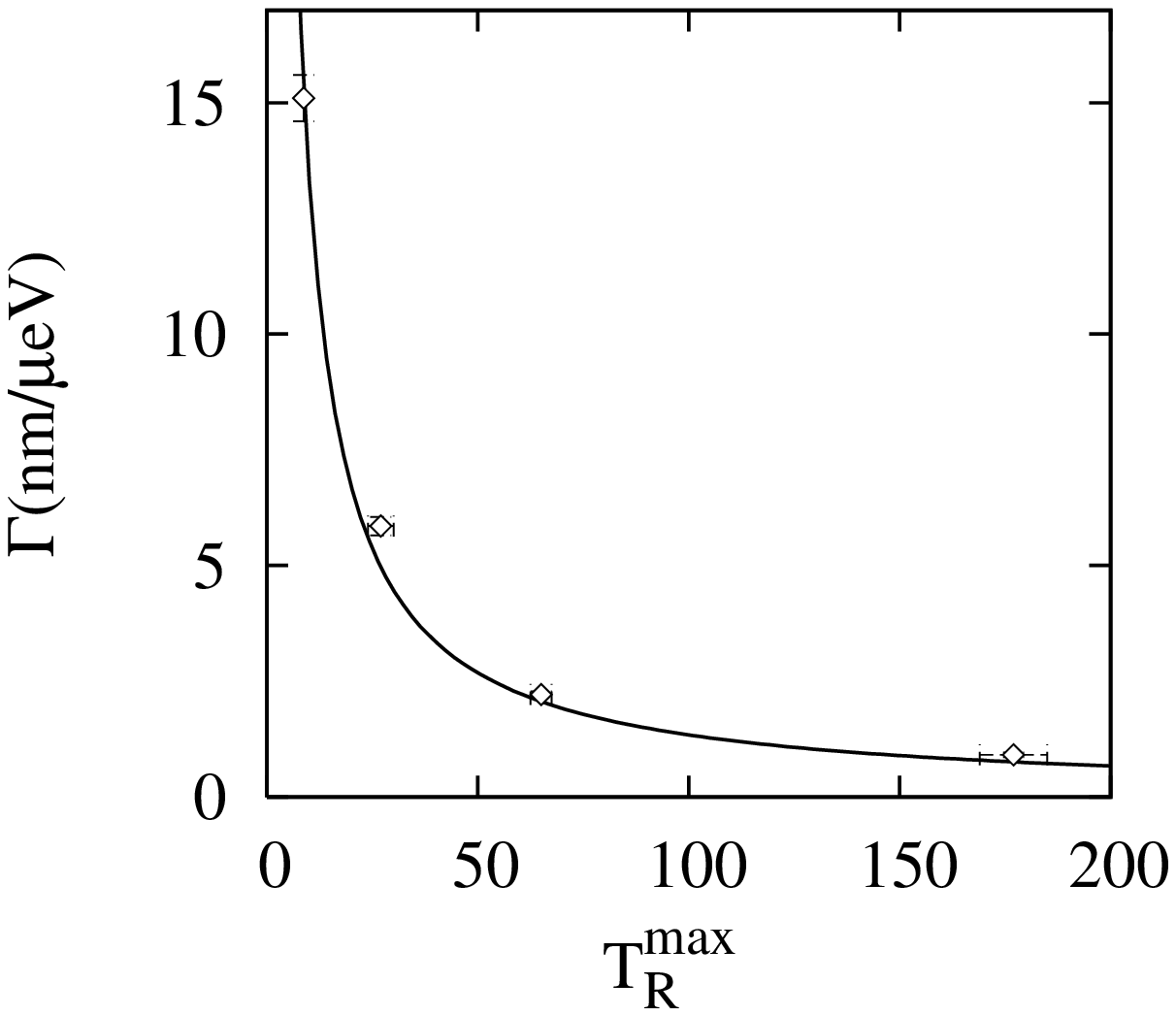}
       \caption[]{Logarithmic fit of $\Gamma_n$ versus resonance index for $\R_{12}(\A)$ (left)
                    and its relation to $T_R^{max}$ (right).}
      \label{fig:04}
      \end{center}
    \end{figure}

    \begin{figure}[b]         %fig05
      \leavevmode
      \begin{center}
        \includegraphics[height=4cm,angle=270,keepaspectratio=true]{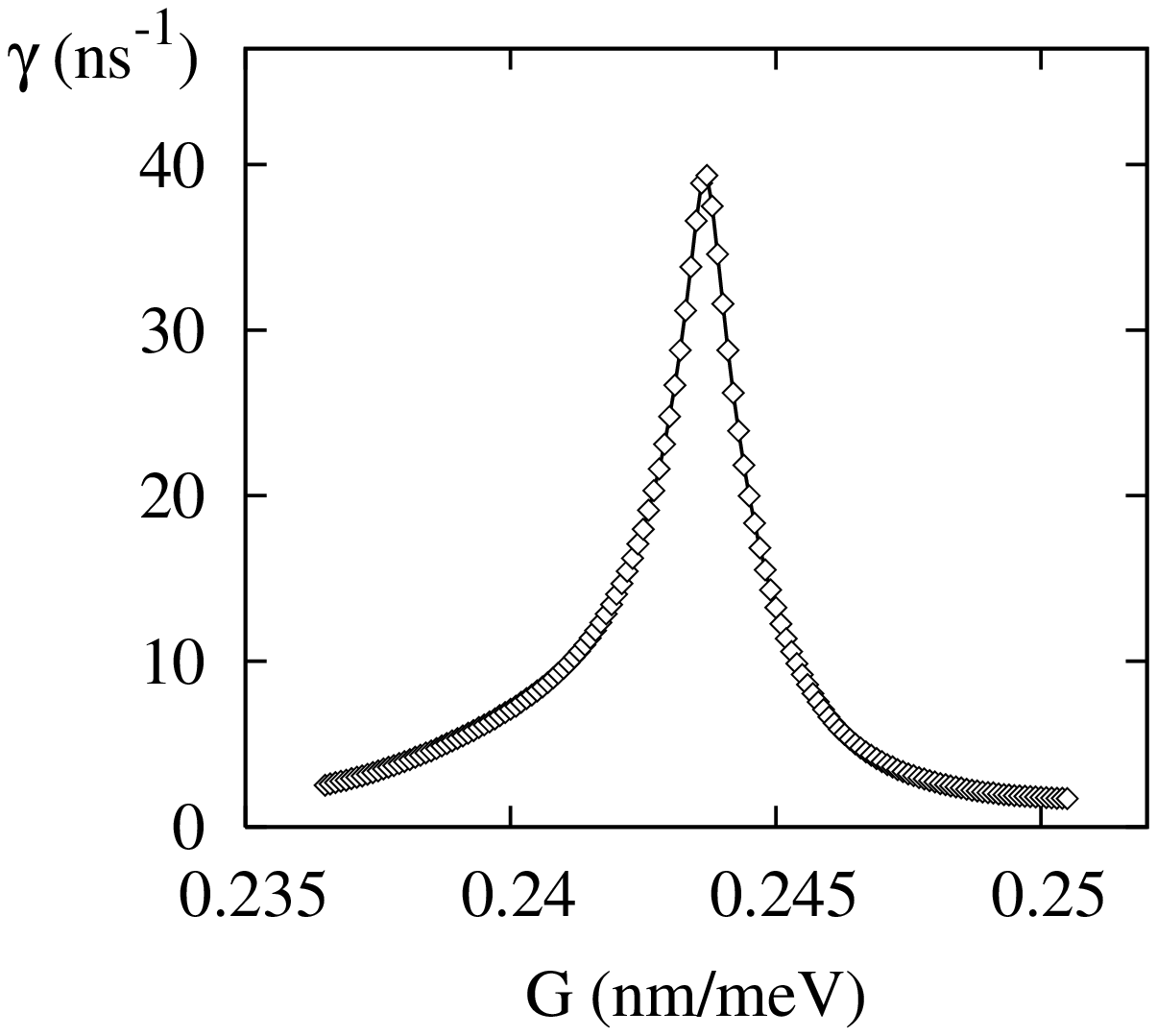}
        \includegraphics[height=4cm,angle=270,keepaspectratio=true]{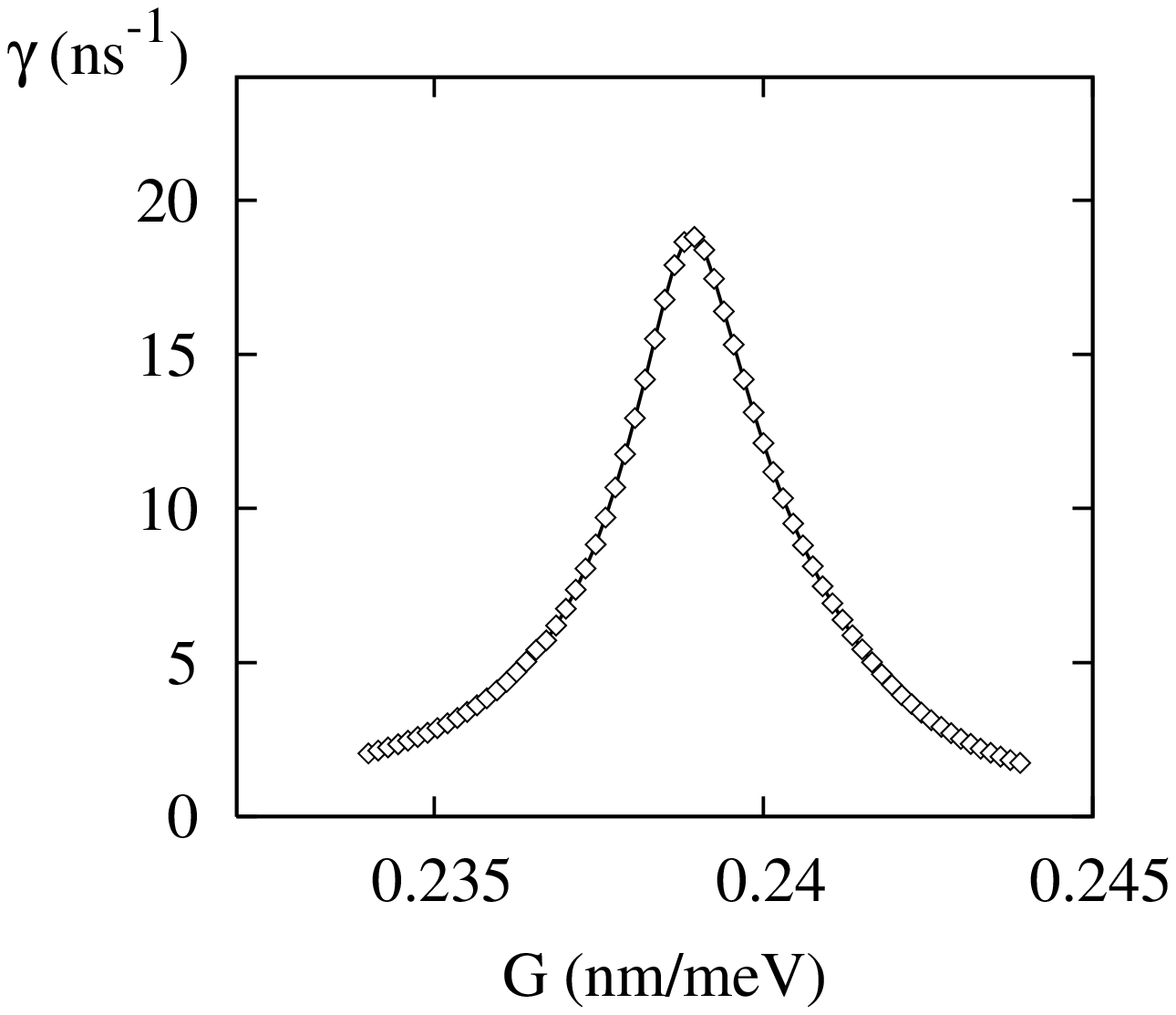}
       \caption[]{Decay rate $\gamma$ of a wavepacket vs. inverse bias, near $\R_{13}^5(\C)$ (left)
                    and $\R_{13}^6(\D)$ (right). }
      \label{fig:05}
      \end{center}
    \end{figure}

    \begin{figure*}         %fig06
      \leavevmode
      \begin{center}
        \begin{tabular*}{1.0\textwidth}{@{\extracolsep{\fill}}ccc}
            \subfigure{\includegraphics[height=5cm,angle=0,keepaspectratio=true]{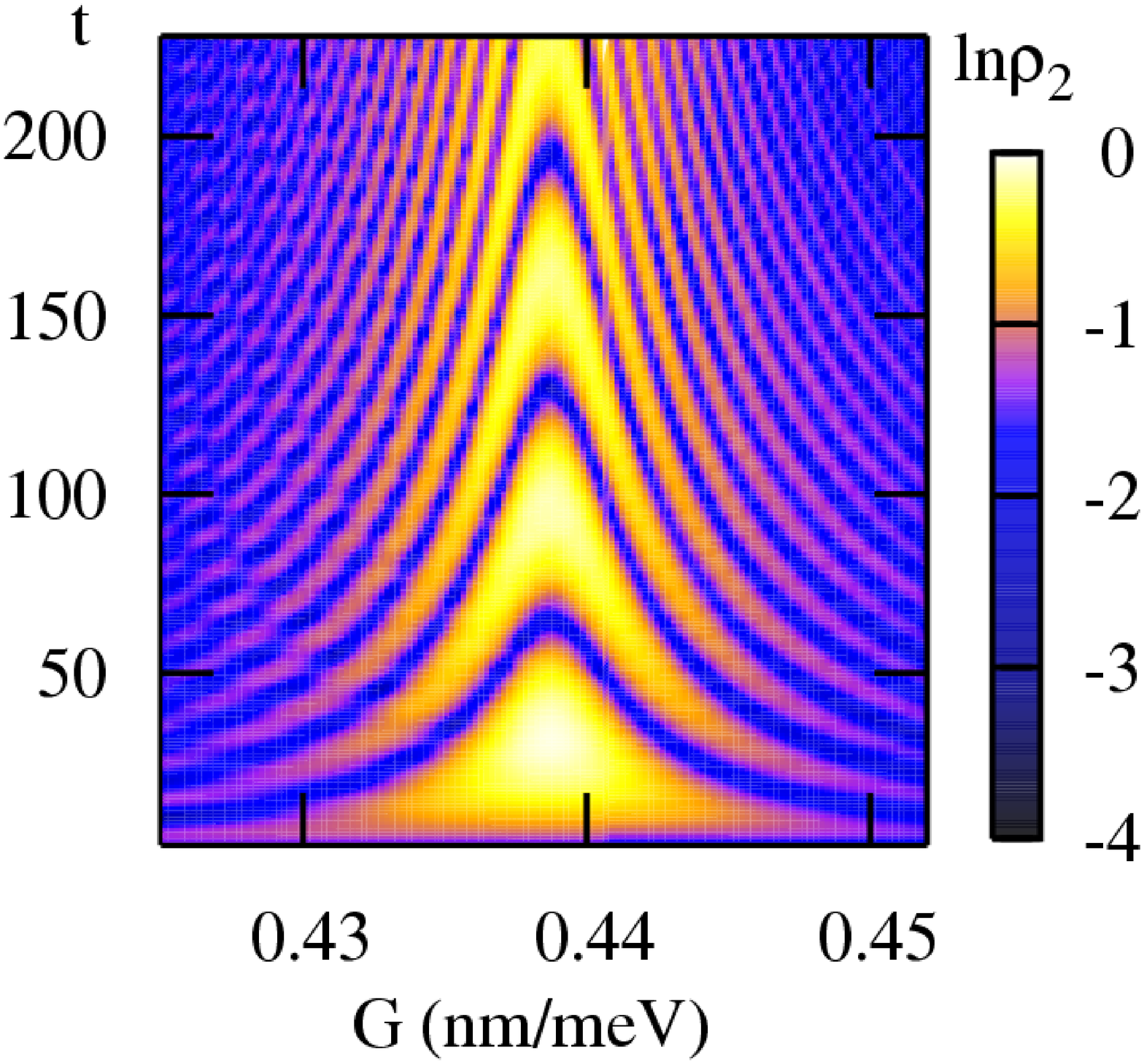}}  &
            \subfigure{\includegraphics[height=5cm,angle=0,keepaspectratio=true]{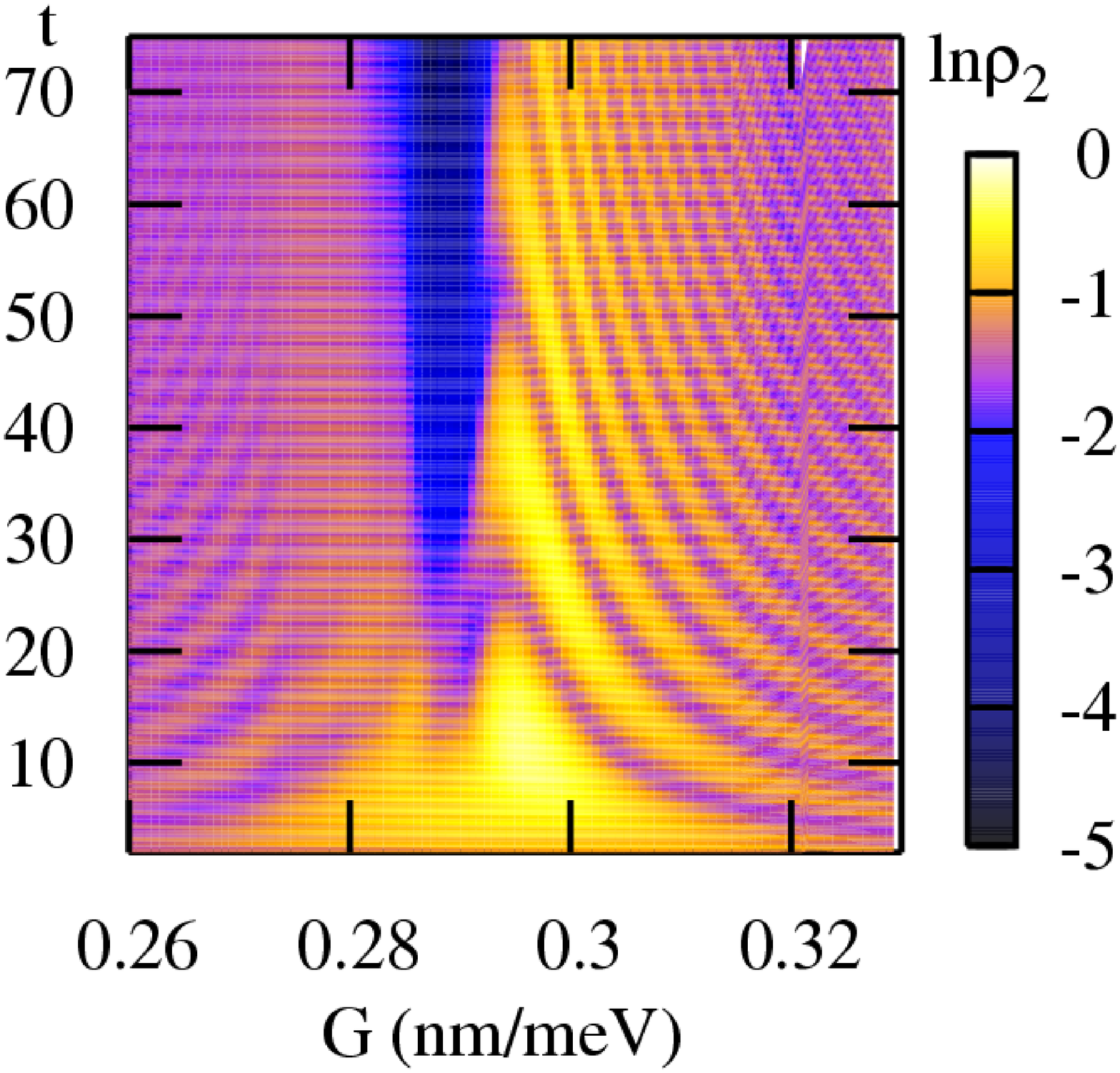}} &
            \subfigure{\includegraphics[height=5cm,angle=0,keepaspectratio=true]{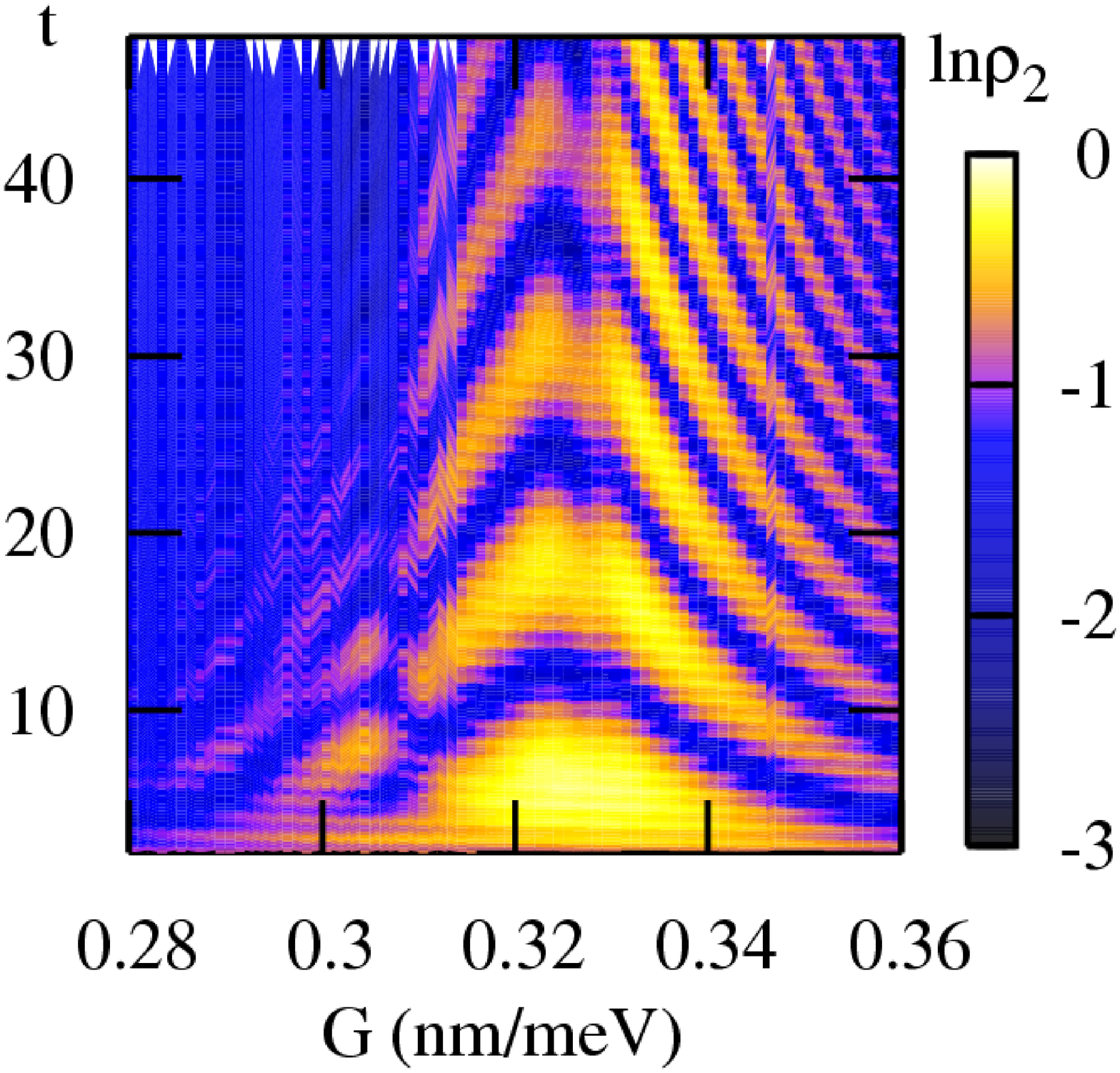}}
        \end{tabular*}
      \end{center}
      \caption[]{Near-resonant dynamics of second miniband occupancy at $\R_{12}^3(\A)$ (left),
                    $\R_{12}^3(\C)$ (center) and $\R_{12}^2(\B)$ (right).}
      \label{fig:06}
    \end{figure*}

As the index $n$ rises, the $\R_{\mu\nu}^n$ are expected to
narrow, since the length of the tunnelling pathway into the
adjacent miniband increases and so does the system's sensitivity to
bias detuning, due to related frequency detuning of the Bloch
oscillations; this is reflected in a decrease of $x_{0n}$ in
Equation~\ref{eq:04}. As predicted by Equation~\ref{eq:04}, the
resonance HWHM dependence on $n$ closely follows an exponential law
(Figure~\ref{fig:04}) with
    \begin{eqnarray}       %eq08
      \Gamma_n &=& \Gamma_1\,\Big( \Gamma_2 / \Gamma_1 \Big)^{n-1}
      \label{eq:08}
    \end{eqnarray}
The anticipated relation
    \begin{eqnarray}       %eq09
      T_n^{max} &=& G_0 /  \Gamma_n
      \label{eq:09}
    \end{eqnarray}
also holds quite well: fitting Equation~(\ref{eq:08}) produced a
value $T_n^{max}\Gamma_n =$$(139~\pm~5)\ \minvunits$ compared to
$G_0=$$(145~\pm~1)\ \minvunits$ for sample A, and calculations for
sample B showed that $T_3^{max}\Gamma_3 =$$(1.05~\pm~0.01)\
\invunits$ compared to $G_0=$$(1.02~\pm~0.01)\ \invunits$.

When an ensemble of resonantly coupled minibands is weakly bound,
Rabi oscillations are weak and overdamped and it is mostly resonant
tunnelling that is seen~\cite{Glutsch04, Zener_recent, Optical2} to
prevail over RO; these represent tunnelling resonances. In this case
the carrier wavepacket escapes to the continuum very quickly and exhibits no
persistent RO. Then $\gamma$ becomes the key parameter
in the dynamical description and features a clear peak centered at
the resonant bias (Figure~\ref{fig:05}). Sensitivity of $\gamma$ to
coupling to higher minibands makes the shape of curves $\gamma(G)$
vary for different resonances.

While we have used perturbation theory
to interpret our calculated results, the good fits obtained by
using Equations~(\ref{eq:08},\ref{eq:09}) suggest that the model
remains valid at strong biases as well. We attribute this to the fact
that the structure of the Wannier-Stark ladder is preserved over the
bias domain considered~\cite{Abumov07}. At
stronger electric fields, however, Bloch oscillations are replaced
by sequential tunnelling, and projection on minibands refers merely to
the wavepacket's distribution in energy (and hence between wells
in real space), rather than a decomposition into Wannier-Stark states.
Indeed, the $n^{th}$ tight-bound miniband contains only harmonics with
wavelengths $\lambda \in [\frac{n}{2}d, \frac{n+1}{2}d]$.

    \begin{figure}[tc]         %fig07
      \leavevmode
      \begin{center}
        \includegraphics[height=4cm,angle=270,keepaspectratio=true]{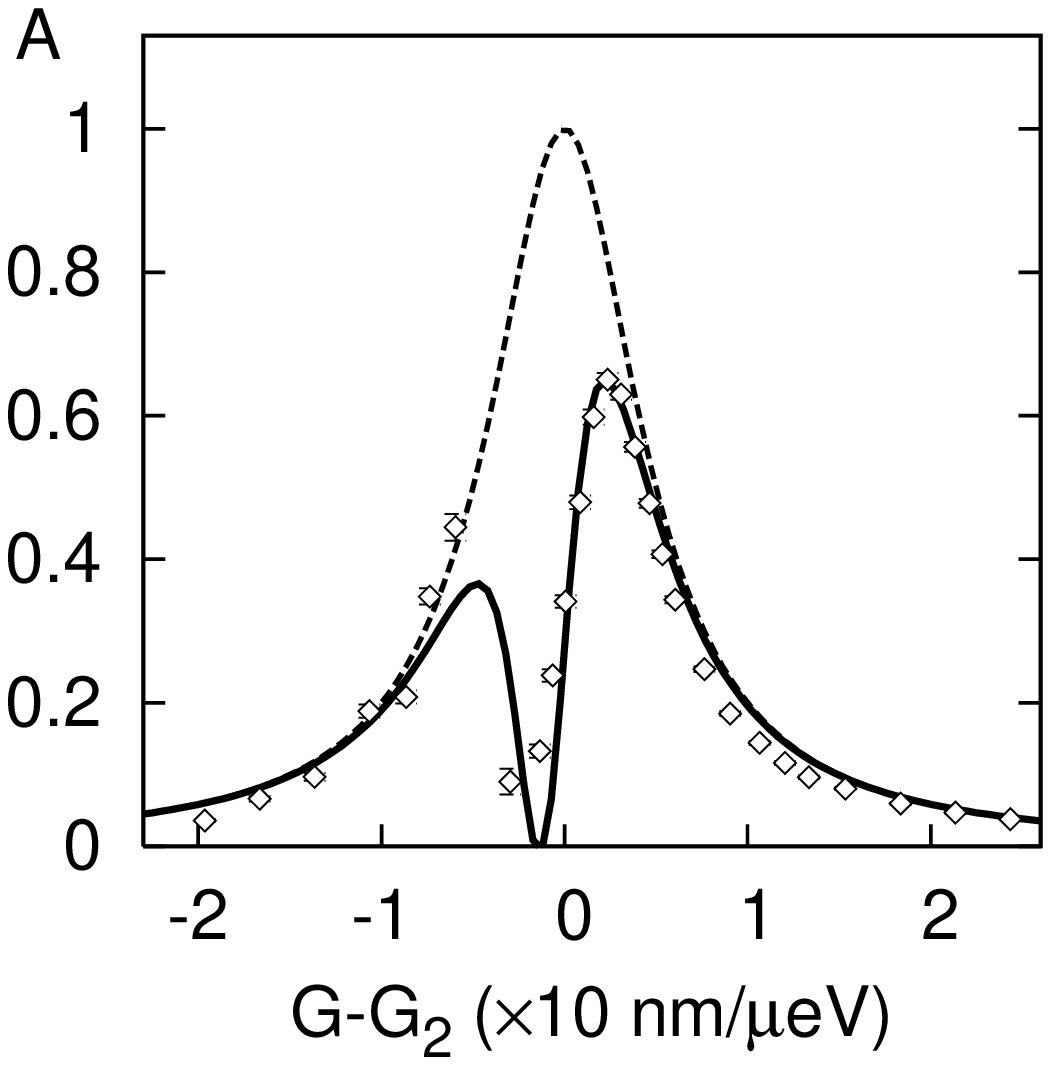}
        \includegraphics[height=4cm,angle=270,keepaspectratio=true]{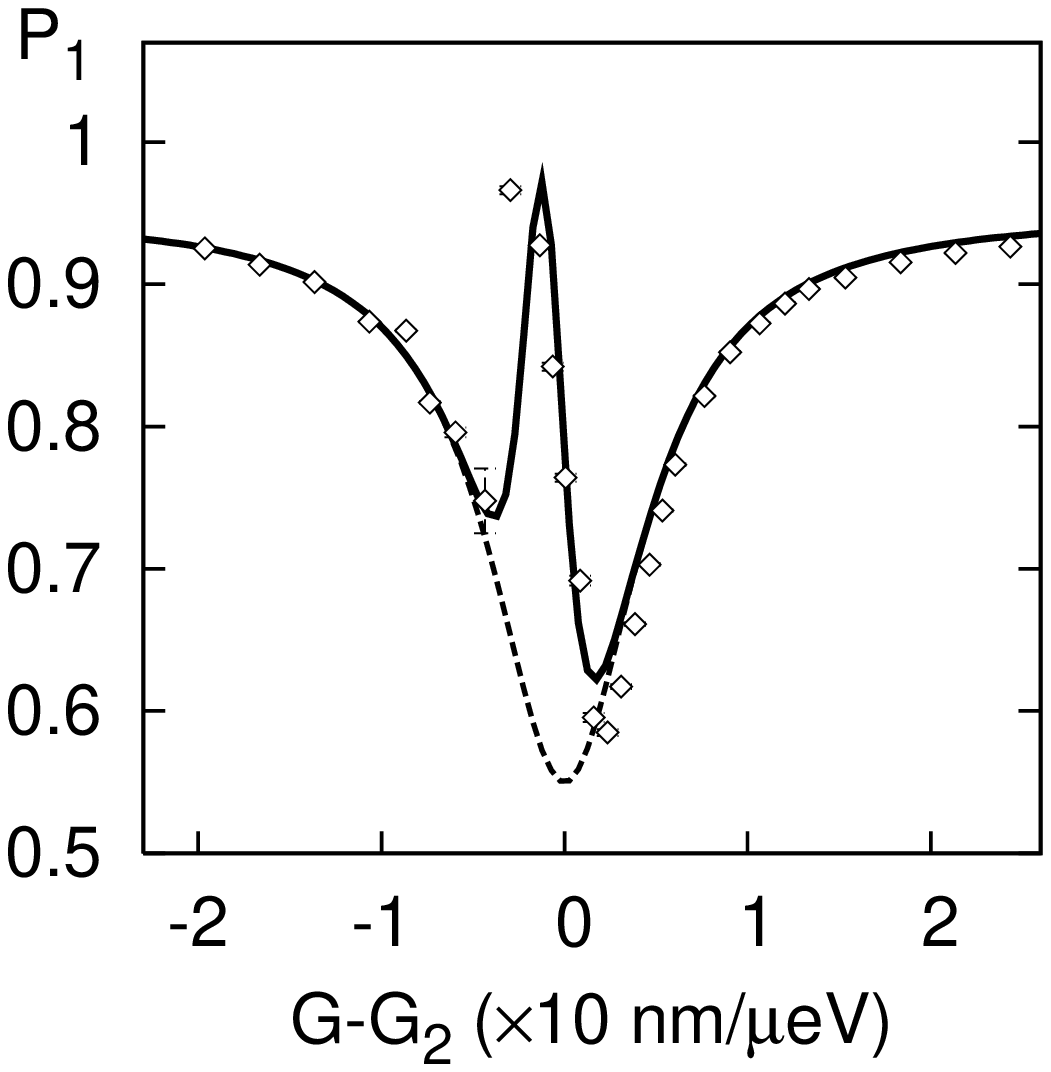}
       \caption[]{Amplitude of RO and of first miniband occupancy near resonances $\R_{12}^2$
                  and $\R_{13}^5$ from middle panel of Figure~\protect\ref{fig:06}.
                  Solid lines show a superposition of Lorentzian curves best fitting simulation data,
                  in circles; broken line is an estimate of a fit in the absence of $\R_{13}^5$.}
      \label{fig:07}
      \end{center}
    \end{figure}

    \begin{figure}[bc]         %fig08
      \leavevmode
      \begin{center}
        \includegraphics[height=4cm,angle=0,keepaspectratio=true]{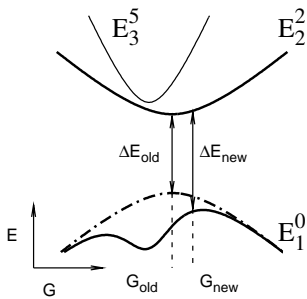}
       \caption[]{Schematic drawing to illustrate the mutual impact of energy level anticrossings.
                 Chain-dotted line shows position of level $E_1^0$ in the
                  absence of repulsion from level $E_3^5$, shown as a thin solid line. For further
                  explanation, see text.}
      \label{fig:08}
      \end{center}
    \end{figure}

\section{\label{sec:RO-RT}Superposition of carrier dynamical patterns}

So far we have considered isolated interminiband resonances. They
occur typically in a strong potential at moderate fields where
resonances have narrow HWHM compared to their spacing in inverse
bias space, and hence they rarely overlap. It is instructive,
however, to consider higher-field resonances in a weaker potential
({\it e.g.} Al$_{0.3}$Ga$_{0.7}$As, sample B in Table 1), where the
interaction between the two resonantly coupled minibands represents
only one of several prominent interference paths. Under these
conditions, the interaction between these paths becomes important
and largely determines the overall dynamics. This gives rise to some
particular dynamical patterns that we interpret as interference of
tunnelling and Rabi resonances, between different pairs of minibands
situated in close proximity.

    \subsection{\label{sec:RO-RT:superpos}Closely situated resonances}

We now consider a case of strongly interfering paths, to illustrate
how different dynamics contribute in such a mixed mode, and gain
understanding of RO stability. This situation typically occurs when
the HWHM of one resonance is large compared to the HWHM of another
resonance that is coupled to the higher minibands. Then one
dynamical pattern is seen to be superimposed on another, as in the
centre panel of Figure~\ref{fig:06} where one can see evidence of
both oscillatory and tunnelling carrier behaviour. The oscillatory
pattern related to the $\R_{12}^2$ resonance, vanishes at the
$\R_{13}^5$ peak, and the $P_1(G)$ and $A_R(G)$ curves in
Figure~\ref{fig:07} demonstrate a sharp extremum resembling a
superposition of two Lorentzian-like curves.

It is worth mentioning that, in the centre panel of
Figure~\ref{fig:06}, the tunnelling resonance $\R_{13}^5$ that
reaches across three minibands, lies above the barrier height of the
SL. This gives it comparable strength to $\R_{12}^2$ acting between
adjacent minibands, which significantly alters the Rabi resonance
pattern. Clearly, resonances which cross three minibands are much weaker
than those that cross just two, and a necessary condition for the former to
affect the latter significantly, is strong bias resulting in weak
carrier localization. Despite such a strong field, we can use a
Wannier-Stark ladder to describe transitions of an electron for the
strong fields considered in Figure~\ref{fig:06}, since the level
structure disrupts only at fields high enough for the potential drop per
unit cell to be close to the smallest interminiband separation in
the system $E_2-E_1$~\cite{Abumov07}. In our case, this value
corresponds to the threshold inverse field $G=0.28\ \invunits$ in
sample B (the strongest potential considered with smallest $E_2-E_1$
separation).

In Figure~\ref{fig:06} we present a few typical cases of
interminiband resonance occurrences. The leftmost panel is an
example of an isolated Rabi resonance having an almost perfectly
symmetric structure; the two panels to the right are examples of
interference between Rabi and tunnelling resonances. In the centre
panel, the tunnelling resonance $\R_{13}^5$ appears to be
particularly strong and dominating. Hence, to the right of the Rabi
resonance $\R_{12}^2$ peak, one sees a clear oscillatory pattern,
whereas to the left of the peak it is smeared out by strong
tunnelling. The close presence of the tunnelling resonance
$\R_{13}^5$ has also the effect of shifting the Rabi resonance
$\R_{12}^2$ peak along the bias scale.

From the relation $F_n=n \, F_0$, the resonance $\R_{12}^2$ (centre
panel of Figure~\ref{fig:06}) should have a peak at
$G=(0.284~\pm~0.002)\ \invunits$; however as a fit to the maximum
period of oscillations indicates, it occurs at a lower bias
$G=(0.295~\pm~0.002)\ \invunits$. The shift occurs because at the
energy level anticrossing corresponding to $\R_{13}^5$, the levels
$E_0^1$ and $E_3^5$ are repelled from each
other~\cite{Gluck_review}. At the same time, the position of level
$E_2^2$ involved in a Rabi resonance remains almost unchanged.
Provided that the minimum mismatch of coupled energy levels ($E_2^2-
E_0^1$) produces the longest period of RO~\cite{Abumov07}, this
results in a shift of the Rabi resonance $\R_{12}^2$ peak as
qualitatively shown in Figure~\ref{fig:08}. There the minimum energy
difference at $\R_{12}^2$, $\Delta E_{old}$, located at $G_{old}$ in
the absence of a $\R_{13}^5$ anticrossing,  shifts to $G_{new}$ due
to superposition of the energy level anticrossings $\R_{12}^2$ and
$\R_{13}^5$. Since the repulsion from the two anticrossings adds
constructively, $\Delta E_{new}$ is larger than $\Delta E_{old}$,
and since the RO period is inversely proportional to the mismatch in
energy level alignment, the peak period of RO shifts to a new value
of bias. Thus the proximity of another resonance has a twofold
effect on the resonance considered: it shifts its resonant bias and
reduces the maximum period of RO. The latter effect is demonstrated
in the right panel of Figure~\ref{fig:06}, where a weak and narrow
tunnelling resonance spread over $G=0.322~\ldots~0.330\ \invunits$
reduces the RO period.

    \subsection{\label{sec:RO-RT:damping}Conditions for Rabi resonance}

    \begin{figure} [b]         %fig09
      \leavevmode
      \begin{center}
        \includegraphics[height=4cm,angle=0,keepaspectratio=true]{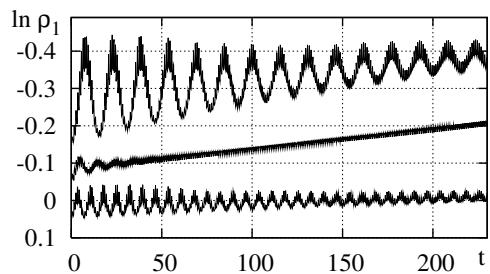}
       \caption[]{Dynamics of first miniband occupancy in the vicinity of $\R_{12}^2$,
                    from the centre panel of Figure~\protect\ref{fig:06}
                     at $G^{(3)}=0.301\ \invunits$ (upper curve), $G^{(2)}=0.293\ \invunits$ (middle
                     curve) and $G^{(1)}=0.274\ \invunits$ (lower curve).
                  The upper [lower] curve is shifted upwards [downwards] by 0.1 for visibility.}
      \label{fig:09}
      \end{center}
    \end{figure}

The simulation data reveal that the decay of RO is slower far from
tunnelling resonances and in stronger potentials, with tightly bound
lowest minibands reducing the carrier decay rate $\gamma$. At the
same time, the RO decay rate nearly vanishes at the very peak of a
Rabi resonance, where the wavepacket decay rate is at its highest.
Thus tunnelling rate alone is not a reliable indicator of Rabi
oscillation damping. As Figure~\ref{fig:09} demonstrates, resonant
tunnelling and Rabi oscillation patterns may occur for the same
wavepacket decay rate.

In the centre panel of Figure~\ref{fig:06} at inverse bias
$G^{(3)}=0.312\ \invunits$ there is a clear RO pattern, whereas at
the bias $G^{(1)}=0.274\ \invunits$ (which is symmetric to $G^{(3)}$
in relation to the Rabi resonance peak located at $G=0.293\ \invunits$), Rabi
oscillations die out quickly. This behaviour goes beyond the
symmetric structure of an isolated resonance as in
Equation~\ref{eq:06} (left panel of Figure~\ref{fig:06}); it is
interference of a tunnelling resonance at $G^{(1)}$ that overdamps
the Rabi oscillations. To generalize, the alignment of energy levels
from higher minibands sets the wavepacket behaviour model and the
link between RO and tunnelling resonance (both being a product of
wavepacket self-interference) is determined by a particular
arrangement of Wannier-Stark ladders: specifically, by the ratio of
interminiband tunnelling rates, as will be argued next.

    \begin{figure}[t]         %fig10
      \leavevmode
      \begin{center}
        \includegraphics[height=4cm,angle=270,keepaspectratio=true]{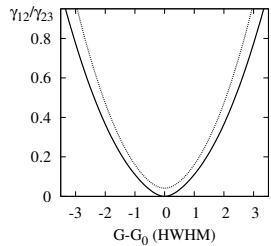}
       \caption[]{Ratio of interminiband tunnelling rates $\gamma_{12}/\gamma_{23}$ vs.
    inverse bias (in resonance HWHM units), from Equation~\protect\ref{eq:10}
    given Equation~\protect\ref{eq:07}. Solid and broken lines show the cases when
    $P_1$ = 0.5 and 0.6 at resonance (ideal and nearly ideal cases), respectively.}
      \label{fig:10}
      \end{center}
    \end{figure}

For illustrative purposes, we turn again to the two-miniband
model and denote the rate of tunnelling from MB$\nu$ to MB$\mu$ as
$\gamma_{\nu\mu}$; this is a convenient measure of strength of
interminiband coupling. The magnitude of $\gamma_{\nu\mu}$ is
inversely proportional to the difference in energy between the
resonantly coupled energy levels from WSL$\nu$ and
WSL$\mu$~\cite{Cohen93}. In the centre panel of
Figure~\ref{fig:06}, these level pairs are: $E_1^0$ and $E_2^2$ for
the Rabi resonance and $E_1^0$ and $E_3^5$ for the tunnelling
resonance. As seen in Figure~\ref{fig:09}, for the inverse biases
$G^{(2)}$ and $G^{(3)}$ featuring the same carrier decay rate, the
difference $(E_2^2-E_1^0)-\Delta E_{21}$ determining the tunnelling
between MB1 and MB2 is -2~meV and 6.2~meV, respectively ($\Delta
E_{21}$ is the energy spacing between the centres of MB2 and MB1,
which equals 88.8~meV for potential A), whereas at the same values
of bias, $(E_3^5-E_1^0)-\Delta E_{31}$ for the ensemble of MB1 and
MB3 equals -10~meV and 10~meV. Considering that the widths of the
tight-binding MB1, MB2 and MB3 are 8.8, 39.4 and 98.6~meV, the
disparity between the two values of $(E_2^2-E_1^0)-\Delta
E_{21}\propto1/\gamma_{12}$ is much more significant than that of
$(E_3^5-E_1^0)-\Delta E_{31}\propto1/\gamma_{13}$. Hence, the ratio
$\gamma_{12}/\gamma_{23}$, which indicates the isolation of
the coupled bands MB1 and MB2 from higher minibands, is much
smaller in the case of $G^{(1)}$. Due to a weaker isolation of the
ensemble at $G^{(1)}$, the quasibound oscillating part of the
wavepacket is more subject to tunnelling into the continuum, so RO
are destroyed and turn more quickly into resonant tunnelling.
The evidence for this is the middle line in
Figure~\ref{fig:09} featuring strongly damped oscillations.

This also explains why, at the peak of an isolated Rabi resonance,
the RO damping rate is the lowest. For example, at $R_{12}$ two
lowest minibands merge through resonant coupling, so $\gamma_{12}$
and $\gamma$ are exploding while $\gamma_{23}$ changes relatively
little. In other words, at an anticrossing, the decay rate of a more
stable state changes much more than that of a less stable one. Thus,
a surging ratio of $\gamma_{12}/\gamma_{23}$ makes RO decay go
nearly to zero at the peak, despite homogeneous level broadening
which reaches its greatest extent there. Away from the peak $\gamma$
falls off and RO persistence decreases, because of the quick drop in
$\gamma_{12}$ (Figure~\ref{fig:03}). This is a somewhat
counterintuitive example of how a quasibound state with shorter
lifetime can produce more prominent oscillations.

Since $\gamma_{12}$ is closely related to the interminiband
transition matrix element $V_{12}$, we can estimate the
near-resonant behaviour of the latter by invoking some dynamical
equilibrium considerations. From an elementary two-miniband model
(see the Appendix) we obtain
    \begin{eqnarray}       %eq10
      \frac{\gamma_{12}}{\gamma_{23}} &=& \frac{2 P_1\,-\,1}{P_1(1\,-\,P_1)}
      \label{eq:10}
    \end{eqnarray}
Neglecting tunnelling pathways between poorly aligned energy levels
({\it e.g.} $E_1^0\rightarrow E_2^1$ at $\R_{12}^2$) and over
many potential wells (in case of resonance across three minibands),
$V_{12}=\gamma_{12}$.  Assuming that $\gamma_{23}$ changes over bias
near $\R_{12}$ adiabatically slowly compared to $\gamma_{12}$, we
get $V_{12} \propto \frac{(2 P_1\,-\,1)}{ P_1(1\,-\,P_1)}$; with the
fit for $ P_1(G)$ from Equation~\ref{eq:07}, we can estimate the
behaviour of $V_{12}$ as shown in Figure~\ref{fig:10}. In the ideal
case of $P_1=0.5$ at the peak, $\gamma_{12}$ vanishes, which
reflects a complete merger of the two lowest minibands. This example
shows a link between the results obtained for bias detuning
dependence of the key dynamical parameters, and theoretical
values. With few simplifying approximations, one can investigate
near-resonant behaviour of theoretical parameters
based on carrier
dynamics in a finite superlattice system in the above manner, using
a system of coupled damped oscillators described in~\cite{Damen90}.

    \subsection{\label{sec:ripples} RO damping and quantum interference}

On a fundamental level, RO damping may be
seen to originate from self-interference of a wavepacket that is
facilitated by interference between small fractions of wavefunction
that a wavepacket constantly emits through intrawell and Bloch
oscillations (we will call them `ripples'). A wavepacket initially
set in the ground miniband of a biassed SL, starts to leak out into
the second miniband, since the interference of the `ripples' is
constructive at the location of the initially unpopulated MB2, and
destructive at MB1. When most of the wavepacket has tunneled out
into MB2, the tunnelling direction reverses. This follows since by
then most of the emission is from MB2. In the ideal case of zero RO
damping, this reversion is a mirror process; in practice, the
tunnelling rate to higher minibands cannot be neglected. This causes
the amount of emitted `ripples' to reduce faster in MB2; such an
inequality breaks the mirror symmetry of the two transitions. The
oscillation pattern smears out and RO are damped more quickly, the
larger is the imbalance in the rate of `ripples' escaping to the
continuum. That is, inversely proportional to
$\gamma_{12}/\gamma_{23}$. In other words, if $\gamma_{23}$ cannot
be neglected compared to $\gamma_{12}$, it breaks the anisotropy in
the system's tunnelling pathways: in addition to
MB1$\leftrightarrow$MB2, there appear two additional pathways,
MB1$\leftrightarrow$MB3 and MB2$\leftrightarrow$MB3 with  very
different tunnelling rates.

\section{\label{sec:conclusion}Conclusion}

From the results of numerical simulations of near-resonant carrier
dynamics at an isolated resonance, we have proposed a set of
equations governing wavepacket behaviour; the dynamical parameters
near a resonance were found to exhibit extrema of various shapes.
For overlapping resonances, a superposition of energy level
anticrossings produces a shift in resonant bias. It also reduces the
maximum period of RO and perturbs the dependence of dynamical
parameters on bias. A superposition of Rabi and tunnelling
resonances also allowed us to observe the transition between
oscillatory and tunnelling coherent wavepacket dynamics, and to
examine the mechanism of RO damping.

Persistence of Rabi oscillations near a resonance was demonstrated
to be independent of homogeneous level broadening and to depend on
the ratio of interminiband tunnelling rates
$\gamma_{12}/\gamma_{23}$, which can serve as an estimate of the RO
damping rate.

\ack
We are grateful to NSERCanada for continuing support under
discovery grant RGPIN-3198. We also thank Dr. W. van Dijk for his
help in implementing the numerical algorithm. The numerical
simulations were carried out on the Shared Hierarchical Academic
Research Computing Network (www.sharcnet.ca).

\section*{Appendix: dynamical equilibrium for a two-miniband system}
\label{app:A}

We consider an elementary system of two strongly interacting
minibands coupled to  continuum states. Further, let us denote the
tunnelling rate from MB$\nu$ to MB$\mu$ as $\gamma_{\nu\mu}$, from
MB$\nu$ to continuum as $\gamma_{\nu\infty}$ and the current
relative occupancy of MB$\nu$ as
$\rho_{\nu}/\rho\equiv\tilde{\rho_{\nu}}$. In steady tunnelling
mode, the miniband occupancies are in dynamical equilibrium.
Neglecting direct tunnelling from MB1 to the continuum, the flow of
probability is balanced as follows:
    \begin{eqnarray}
        \left\{\begin{array}{ll}
            \frac{d}{dt}\vec{\tilde{\rho}} \,=\,
                \left( \begin{array}{cc}
                    \gamma_{12} \ &  -\gamma_{21}-\gamma_{2\infty} \\
                    -\gamma_{12} \ &   \gamma_{21}
                \end{array} \right)
                \vec{\tilde{\rho}}       & \textrm{} \\

               \frac{d}{dt}\frac{\tilde{\rho_1}}{\tilde{\rho_1}+\tilde{\rho_2}}  \,=\,
               - \frac{d}{dt}\frac{\tilde{\rho_2}}{\tilde{\rho_1}+\tilde{\rho_2}}    & \textrm{}
        \end{array}\right.  \nonumber
    \end{eqnarray}
where $\vec{\tilde{\rho}}$ has components $\tilde{\rho}_1,\,
\tilde{\rho}_2$, and the second equation expresses the condition for
dynamical equilibrium. A little algebra produces the quadratic
equation
    \begin{eqnarray}
        \kappa^2 \gamma_{12} - \kappa (\gamma_{21} - \gamma_{12} + \gamma_{2\infty} ) - \gamma_{21} &=& 0 \nonumber
    \end{eqnarray}
with $\kappa={\tilde{\rho_1}}/{\tilde{\rho_2}}>1$. Taking
$\tilde{\rho_2}=1-\tilde{\rho_1}$ and assuming
$\gamma_{12}\approx\gamma_{21}=\gamma$ (which holds well in the
vicinity of a resonance peak), in the steady-state limit
$t\rightarrow\infty$ we arrive at
    \begin{eqnarray}
        \alpha &=& \frac{2P_1-1}{P_1(1-P_1)} \nonumber
    \end{eqnarray}
with the notation $\alpha=\gamma_{\infty}/2\gamma$ and
$\gamma_{\infty}=\gamma_{2\infty}$. Note that the extreme case of RT
corresponds to $\alpha\rightarrow\infty$ with $P_1\rightarrow 1$,
and of RO to $\alpha\rightarrow 0$ with $P_1\rightarrow 0.5$.

%#############################################################################################

\section*{References}

%\bibliographystyle{nar}
%\bibliography{references}

\end{document}